\documentclass[aps,prapplied,twocolumn,superscriptaddress,amsmath,amssymb]{revtex4-2}

\usepackage{graphicx}
\usepackage{bm}
\usepackage{hyperref}

\begin{document}

\title{Valley-Landscape Engineering in Bilayer WSe$_2$ Gate-All-Around
  Transistors}

\author{Katsunori Wakabayashi}
\email{WAKABAYASHI.Katsunori@nims.go.jp}
\affiliation{
  Research Center for Materials Nanoarchitectonics (MANA),
  National Institute for Materials Science (NIMS),
  Tsukuba 305-0044, Japan}
\affiliation{Kwansei Gakuin University, Sanda 669-1330, Japan}

\author{Souren Adhikary}
\affiliation{
  Research Center for Materials Nanoarchitectonics (MANA),
  National Institute for Materials Science (NIMS),
  Tsukuba 305-0044, Japan}

\author{Kazuhito Tsukagoshi}
\affiliation{
  Research Center for Materials Nanoarchitectonics (MANA),
  National Institute for Materials Science (NIMS),
  Tsukuba 305-0044, Japan}

\date{\today}

\begin{abstract}
The K--$\Gamma$ valence-band splitting $\Delta_{K\Gamma}$ that governs
hole transport in few-layer WSe$_2$ is not a fixed material constant.
It is tuned by interlayer stacking, strain, pressure, and the dielectric
and displacement-field environment.
Its computed value also depends on the level of electronic-structure
theory.
Bilayer WSe$_2$ therefore realizes a tunable multivalley landscape
rather than a single operating point.
We combine first-principles inputs with an analytical two-valley device
model for gate-all-around (GAA) field-effect transistors, and obtain
three results.
(i)~The minimum subthreshold swing stays near the thermionic limit of
$60$~mV~dec$^{-1}$ independently of layer number, because well below
threshold the quantum capacitance remains far below the oxide
capacitance.
(ii)~The effective mobility is set by the K-to-$\Gamma$ occupation ratio:
valley redistribution is strong when $\Delta_{K\Gamma}$ is of order
$k_BT$ and fades toward single-valley $K$ transport as the splitting
grows.
(iii)~In this small-splitting regime, biaxial strain tunes the effective
mobility---and hence the on-current---through the valley population while
the subthreshold swing stays at the thermionic limit, decoupling mobility
control from electrostatic switching in a way distinct from
scattering-based strategies.
A symmetric GAA gate controls carrier density at essentially fixed
splitting, and because an out-of-plane field \emph{increases}
$\Delta_{K\Gamma}$, its midplane-symmetric potential avoids driving the
channel out of the small-splitting regime.
The design principle has two facets: small $\Delta_{K\Gamma}$---set by
layer number, stacking, strain, and dielectric engineering---maximizes
the valley \emph{tunability} of $\mu_\mathrm{eff}$, whereas larger
$\Delta_{K\Gamma}$, for example through compressive strain, suppresses
the heavy $\Gamma$ valley and maximizes the on-state \emph{mobility}.
\end{abstract}

\maketitle

\section{Introduction}

The valley degree of freedom in two-dimensional (2D) transition metal
dichalcogenides (TMDCs) has attracted broad interest as a platform for valleytronic
devices\cite{Xiao2012_valley,Manzeli2017_TMDreview,Novoselov2016_2Dscience,Geim2013_vdW,Lai2023_valleytronic}.
Among TMDCs, WSe$_2$ exhibits a particularly rich valence-band
structure. The valence-band maximum (VBM) resides at the $K$ point in
the monolayer but shifts toward the $\Gamma$ point as the layer number
increases, driven by interlayer coupling and spin--orbit
interaction\cite{Zhao2013_WSe2layers,Zhang2014_WSe2ARPES,Wickramaratne2014_indirect,Yeh2015_WSe2muARPES,Kormanyos2015_kp,Ridolfi2015,Zhu2011_SOC}.
The K--$\Gamma$ valley splitting
$\Delta_{K\Gamma} = E_v(K) - E_v(\Gamma)$ is large and positive in the
monolayer ($\approx 0.5$~eV), small in the bilayer, and negative in the
trilayer, where $\Gamma$ is the VBM throughout.
The bilayer therefore sits close to the K--$\Gamma$ crossover.
Its precise value, however, is not a fixed material constant.
It is small on the scale of electronic-structure uncertainties and, as we
show below, depends strongly on the exchange--correlation functional,
the van der Waals treatment, and the interlayer stacking.
Experimentally relevant values span a few tens to about $100$~meV,
although different theoretical approximations cover a substantially wider
range.

This layer-number evolution is characterized by density-functional-theory
(DFT) calculations that
include spin--orbit coupling (SOC) and by angle-resolved photoemission
spectroscopy (ARPES)
measurements\cite{Wickramaratne2014_indirect,Zhang2014_WSe2ARPES,Yeh2015_WSe2muARPES}.
In particular, submicron-ARPES on few-layer WSe$_2$ measures the
bilayer (2H) splitting directly as
$E_v(K)-E_v(\Gamma) = 0.04 \pm 0.03$~eV, i.e.\ $K$ remains the
valence-band maximum by ${\approx}40$~meV, with the crossover to a
$\Gamma$-point maximum occurring only in the
bulk\cite{Wilson2017_muARPES}.
Shubnikov--de Haas (SdH) oscillations in hexagonal boron nitride
(h-BN)-encapsulated trilayer
WSe$_2$ have directly confirmed simultaneous K- and $\Gamma$-valley
hole populations, with experimental effective masses
$m^*_K = 0.5\,m_0$ and $m^*_\Gamma = 1.2\,m_0$, where $m_0$ is the
free-electron mass,
and demonstrated that a transverse electric field transfers holes from
the $\Gamma$ valley to the $K$ valley\cite{Movva2018_valleyPRL}.
Yet a quantitative framework connecting this valley structure to
room-temperature field-effect-transistor (FET) characteristics---subthreshold swing (SS), effective
mobility, drain current---in the technologically relevant bilayer
limit has been lacking.

The physical significance of the near-degeneracy in the bilayer is
that the $K$ and $\Gamma$ valleys differ markedly in hole effective
mass, with $m^*_\Gamma \gg m^*_K$ across all strain
values\cite{Wickramaratne2014_indirect,Jin2014,Xu2014_spinpseudo,Movva2018_valleyPRL}.
Any redistribution of holes between the two valleys therefore directly
modifies the effective mobility---an effect that becomes pronounced in
the small-splitting regime, where $\Delta_{K\Gamma}$ is of order $k_BT$.
In the monolayer, the large splitting confines holes entirely to the
light $K$ valley, so mobility is governed by single-valley scattering.
In the bilayer, gate-induced carrier accumulation can drive a
continuous shift of the hole population from $K$ toward $\Gamma$,
producing a gate-tunable effective mobility with no direct counterpart in
conventional single-valley materials.
This valley-redistribution mechanism constitutes a distinct transport
regime. It is an intrinsic band-structure effect, not an extrinsic
scattering effect, and it is most pronounced near the bilayer
K--$\Gamma$ degeneracy.

WSe$_2$ has emerged as a leading p-channel candidate for
field-effect transistors, with extensive experimental
characterization of mobility, subthreshold swing, and contact
resistance\cite{Fang2012_WSe2pFET,Liu2013_WSe2nFET,Allain2014_WSe2mobility,Pradhan2015_WSe2Hall,Smets2023_WSe2improvements,Chiu2025VLSI,Ghosh2025_bilayerWSe2FET}.
In particular, large-area bilayer WSe$_2$ films with low defect
density---grown by alkali-assisted chemical vapor deposition---have
recently been shown to support hole mobilities near
$120~\mathrm{cm^2\,V^{-1}\,s^{-1}}$ and reduced contact resistance
relative to the monolayer, underscoring the technological relevance of
the bilayer channel studied here\cite{Chou2025_bilayerWSe2growth}.
Among device geometries, the gate-all-around (GAA) architecture
provides the strongest electrostatic control, as the gate electrode
fully surrounds the channel on all
sides\cite{Loubet2017_nanosheet,Ionescu2011,Mukesh2022_GAA_review,Desai2016}.
Unlike a planar FET, the symmetric wrap-around gate also suppresses the
antisymmetric interlayer electric field responsible for the Stark
renormalization of the valley splitting, so that the gate tunes the
carrier density while leaving the valley landscape essentially intact.

Atomically thin channels are especially well suited to this geometry
owing to their sub-nanometer thickness and van der Waals
interfaces\cite{Das2021_2Dreview,Fiori2014_2Delectronics,Pan2025_TMDscaling,Huang2024_TMDlimits}.
Recent integration of ferroelectric high-$\kappa$ dielectrics such as
Hf$_{0.5}$Zr$_{0.5}$O$_2$ (HZO) in GAA structures confirms the
feasibility of strong vertical electric-field control in 2D-channel
devices\cite{Muller2011APL,Lin2025NatElectron}, while the interplay
between the gate field and the interlayer Stark effect is examined in
Sec.~\ref{sec:discussion_efield}.
Biaxial strain, which arises naturally from substrate mismatch or can
be applied
intentionally,\cite{He2013,Conley2013,Johari2012,Scalise2014,He2024_strainFET,Schmidt2024_straintransfer,Ahn2017NatComm,Carrascoso2021_strainTMD,Jaikissoon2024_strainCMOS}
provides a complementary tuning parameter. It continuously tunes $\Delta_{K\Gamma}$
through modification of the interlayer coupling.
Despite these advances, the framework identified above---one that
links the first-principles valley structure of few-layer WSe$_2$ to
measurable device characteristics---has yet to be established.

In this work, DFT calculations with spin--orbit coupling and van der
Waals corrections are combined with an analytical two-valley device
model to address this need.
Because the bilayer splitting is method- and structure-dependent, we
treat $\Delta_{K\Gamma}$ as a tunable landscape parameter---set
physically by stacking, strain, pressure, and the dielectric and electrostatic environment
(and, in theory, by the exchange--correlation functional)---and map how
it controls transport, rather than relying on a single computed value.
The key results are the following.
(i)~The minimum subthreshold swing remains near the thermionic limit of
$60$~mV~dec$^{-1}$ independently of layer number, because sufficiently
below threshold the quantum capacitance remains well below the oxide
capacitance.
(ii)~The effective mobility is governed by the K-to-$\Gamma$ occupation
ratio, and the valley-redistribution mechanism is strong in the
small-splitting regime ($\Delta_{K\Gamma}$ of order $k_BT$), fading toward
single-valley $K$ transport as the splitting grows.
(iii)~In this small-splitting regime, biaxial strain tunes
$\mu_\mathrm{eff}$ and the on-current through the valley population
while SS remains near the thermionic limit; because the off-state
current scales with the same $\mu_\mathrm{eff}$ at fixed equilibrium
density, the on/off ratio is left nearly strain-independent.
This decoupling of mobility control from the subthreshold swing arises
because valley redistribution shifts holes between valleys of different
transport mobility, without altering the electrostatic gate efficiency.
In a symmetric gate-all-around geometry the gate controls carrier
density at essentially fixed splitting, because the antisymmetric
interlayer field that would otherwise modify $\Delta_{K\Gamma}$ through
the Stark effect vanishes to leading order in the ideal
midplane-symmetric limit. Strain, not the gate field, tunes
$\Delta_{K\Gamma}$ (Sec.~\ref{sec:discussion_efield}).

\begin{figure*}[tbh]
  \includegraphics[width=\textwidth]{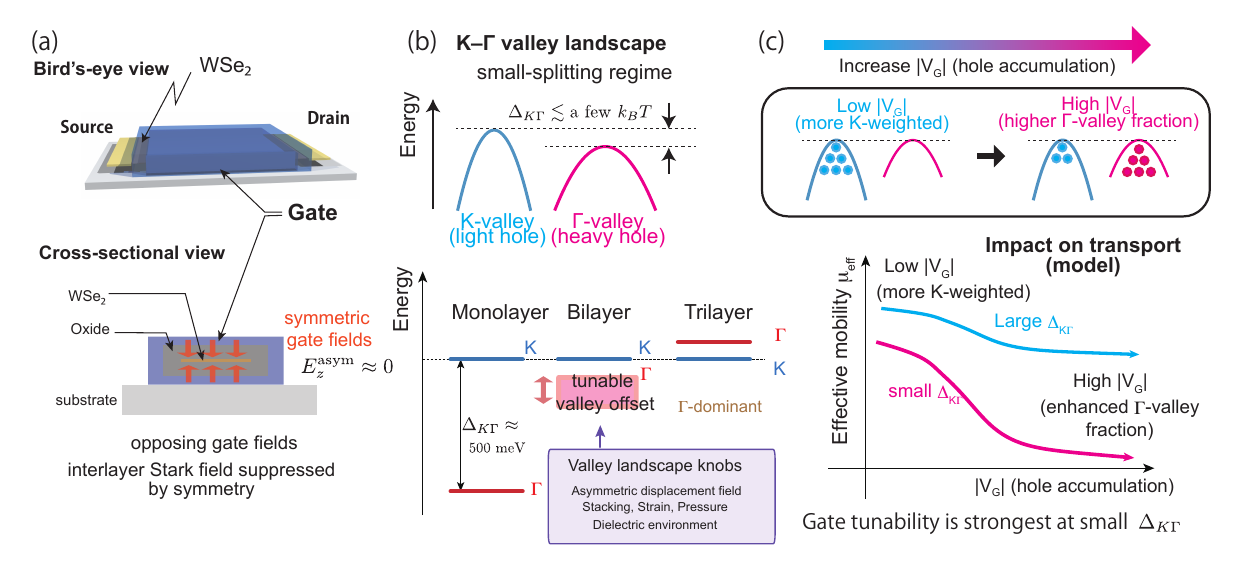}
  \caption{WSe$_2$ GAA FET concept and the K--$\Gamma$ valley landscape.
    (a)~Gate-all-around (GAA) device geometry.
    Top: bird's-eye view of the WSe$_2$ nanosheet channel with source,
    drain, and gate electrodes.
    Bottom: cross-sectional view. The symmetric wrap-around gate produces
    a midplane-symmetric electrostatic potential; the oppositely directed
    vertical fields generate no leading-order interlayer potential
    difference, so the antisymmetric Stark component is suppressed
    ($E_z^\mathrm{asym}\approx0$) and the gate acts as a common-mode
    carrier-density control (Sec.~\ref{sec:discussion_efield}). The
    cross-section is a schematic of the idealized symmetric
    electrostatics; substrate- and dielectric-induced asymmetries are
    neglected.
    (b)~The K--$\Gamma$ valley \emph{landscape}.
    Top: in the small-splitting regime ($\Delta_{K\Gamma}\lesssim$ a few
    $k_BT$) the light $K$-valley and heavy $\Gamma$-valley band edges lie
    within a few $k_BT$, so both are thermally populated.
    Bottom: schematic layer-number evolution---the splitting is large in
    the monolayer ($\Delta_{K\Gamma}\approx500$~meV, $K$ the VBM), small
    and tunable in the bilayer, and $\Gamma$-dominated in the trilayer.
    The realized bilayer offset is a landscape quantity set by the
    asymmetric displacement field, stacking, strain, and dielectric
    environment, and---as established in prior first-principles
    studies---by hydrostatic pressure (Sec.~\ref{sec:robustness}).
    (c)~Impact on transport (model).
    Increasing $|V_\mathrm{G}|$ accumulates holes and progressively
    populates the heavier $\Gamma$ valley (top).
    The effective hole mobility $\mu_\mathrm{eff}$ then decreases with
    $|V_\mathrm{G}|$; the modulation is strong for small
    $\Delta_{K\Gamma}$ and weak for large $\Delta_{K\Gamma}$, where
    transport is single-valley $K$ (bottom).
    Gate tunability of $\mu_\mathrm{eff}$ is thus strongest near the
    K--$\Gamma$ crossover and fades as the splitting grows.}
  \label{fig:device}
\end{figure*}

\begin{figure*}[tbh]
  \includegraphics[width=\textwidth]{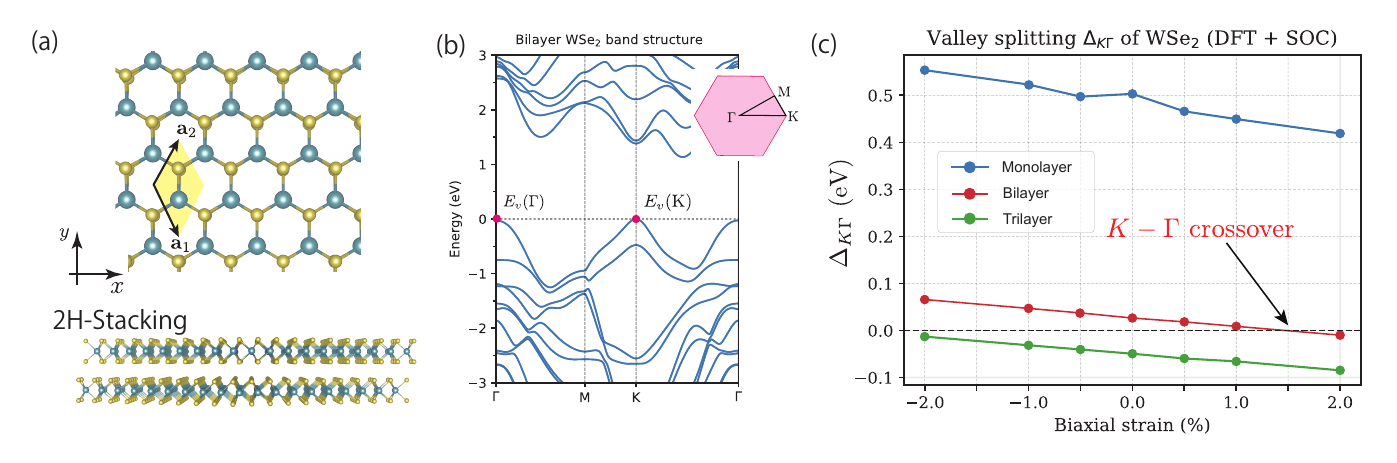}
  \caption{Crystal structure and electronic structure of WSe$_2$.
    (a)~Crystal structure of WSe$_2$.
    Top: top view of the hexagonal lattice with W (blue) and Se
    (yellow) atoms; $\mathbf{a}_1 = (a/2, -a\sqrt{3}/2)$ and
    $\mathbf{a}_2 = (a/2, a\sqrt{3}/2)$ are the primitive lattice
    vectors, $a$ is the lattice constant, and the rhombus indicates
    the unit cell.
    Bottom: side view of bilayer WSe$_2$ in the 2H stacking
    configuration.
    (b)~DFT+SOC band structure of bilayer WSe$_2$ along the
    $\Gamma$--M--K--$\Gamma$ path, computed with the
    generalized-gradient-approximation (GGA-PBE)
    functional including spin--orbit coupling and van der Waals
    corrections.
    The valence-band maxima at $\Gamma$ and $K$ both lie near zero
    energy, giving a small splitting
    $\Delta_{K\Gamma} \approx 0.027$~eV at this (PBE) level of theory;
    the value is method-dependent (Sec.~\ref{sec:robustness}).
    Inset: first Brillouin zone with high-symmetry points.
    (c)~Valley splitting $\Delta_{K\Gamma} = E_v(K) - E_v(\Gamma)$
    as a function of biaxial strain for monolayer, bilayer, and
    trilayer WSe$_2$ from DFT+SOC calculations.
    The horizontal dashed line marks $\Delta_{K\Gamma} = 0$.
    The bilayer undergoes a K--$\Gamma$ crossover near $+1.5\%$
    tensile strain. The trilayer has $\Delta_{K\Gamma} < 0$ at all
    strains studied ($\Gamma$ is always the VBM).}
  \label{fig:crystal}
\end{figure*}

\section{Electronic Structure: DFT Results}
\label{sec:dft}

Figure~\ref{fig:device} summarizes the WSe$_2$ GAA FET concept and
the K--$\Gamma$ valley landscape.
Figure~\ref{fig:device}(a) shows the gate-all-around device geometry,
in which the WSe$_2$ nanosheet channel is fully enclosed by the gate
dielectric and electrode; the symmetric wrap-around gate suppresses the
antisymmetric interlayer field, so the gate controls carrier density at
essentially fixed splitting (Sec.~\ref{sec:discussion_efield}).
Figure~\ref{fig:device}(b) shows the valley landscape: the light $K$ and
heavy $\Gamma$ band edges lie within a few $k_BT$ in the bilayer, and the
splitting evolves from $K$-dominated in the monolayer
($\Delta_{K\Gamma} \approx 500$~meV) to $\Gamma$-dominated in the
trilayer, with the bilayer offset tunable through the landscape
parameters.
Figure~\ref{fig:device}(c) shows the consequence for transport: increasing
gate bias populates the heavier $\Gamma$ valley, so $\mu_\mathrm{eff}$
becomes gate-controllable---most strongly when $\Delta_{K\Gamma}$ is small
(a few $k_BT$) and weakly once the splitting grows.

The crystal and electronic structure underlying this behavior are shown
in Fig.~\ref{fig:crystal}.
Figure~\ref{fig:crystal}(a) shows the crystal structure of WSe$_2$,
where $\mathbf{a}_1 = (a/2, -a\sqrt{3}/2)$ and
$\mathbf{a}_2 = (a/2, a\sqrt{3}/2)$ are the primitive lattice vectors,
$a$ is the lattice constant, and the rhombus indicates the unit cell.
As shown in Fig.~\ref{fig:crystal}(b), at the PBE-D3(BJ)+SOC level the
unstrained bilayer lies in the near-degenerate regime, with the
valence-band maxima at $K$ and $\Gamma$ separated by only
$\Delta_{K\Gamma} \approx k_BT$ (inset: first Brillouin zone).
This PBE result motivates the two-valley analysis. Sec.~\ref{sec:robustness}
then generalizes the model over the broader $\Delta_{K\Gamma}$ landscape,
since the splitting is method- and structure-dependent.
As summarized in Fig.~\ref{fig:crystal}(c), biaxial strain provides a
controlled means to tune $\Delta_{K\Gamma}$ across different layer
numbers, allowing systematic isolation of the valley-population effect
on transport.
Accordingly, $\Delta_{K\Gamma}$ and the hole effective masses for
monolayer, bilayer, and trilayer WSe$_2$ are computed as functions of
biaxial strain. These are the key DFT inputs to the device model of
Sec.~\ref{sec:model}.

\subsection{Computational Details}

All calculations were performed using the Vienna \textit{Ab initio}
Simulation Package (VASP)\cite{Kresse1996PRB,VASP} within the
generalized gradient approximation (GGA-PBE)\cite{PBE} with
spin--orbit coupling.
Van der Waals corrections (DFT-D3(BJ))\cite{DFT_D3} were included for
bilayer and trilayer structures.
Free-standing monolayer, bilayer (2H), and trilayer (2H) WSe$_2$ were
modeled with vacuum spacings of 20, 25, and 35~\AA, respectively.
A plane-wave cutoff of 600~eV was used throughout.
Structural relaxations employed a $\Gamma$-centered $12\times12\times1$
$k$-point mesh with energy and force convergence thresholds of
$10^{-8}$~eV and $10^{-3}$~eV~\AA$^{-1}$, respectively.
Band structures were computed on a denser $32\times32\times1$ mesh.
Biaxial strain $\varepsilon$ was applied by scaling the lattice constant
$a \to a(1+\varepsilon)$ with positions relaxed at each of seven values:
$\varepsilon = -2, -1, -0.5, 0, +0.5, +1, +2$\,\%.
The lattice constant $a$ of monolayer WSe$_2$ without strain is
3.32~\AA, while those of bilayer and trilayer WSe$_2$ in 2H stacking
are 3.28~\AA.

\subsection{Valley Splitting $\Delta_{K\Gamma}$}

Figure~\ref{fig:crystal}(c) shows $\Delta_{K\Gamma}$ as a function of
biaxial strain for monolayer, bilayer, and trilayer WSe$_2$.
The monolayer maintains a large positive $\Delta_{K\Gamma} \approx
0.50$~eV across the entire strain range, confirming robust $K$-valley
dominance.
In the bilayer, $\Delta_{K\Gamma}$ decreases from $0.0658$~eV at
$-2\%$ compressive strain to $-0.0099$~eV at $+2\%$ tensile strain,
passing through a K--$\Gamma$ crossover at $+1.5\%$ by linear
interpolation between the calculated points.
The trilayer has $\Delta_{K\Gamma} < 0$ throughout, with the $\Gamma$
point as the VBM at all strain values
considered\cite{Komsa2012_indirect,Peng2019_WSe2biaxial}.

These results establish a clear hierarchy. Increasing layer number
progressively stabilizes the $\Gamma$-valley VBM, and biaxial tensile
strain further promotes $\Gamma$ occupation.
Among the three, the bilayer is the layer number whose splitting falls
in the small-$\Delta_{K\Gamma}$ range where the two valleys compete. At
the present level of theory the PBE value is
$\Delta_{K\Gamma} \approx 0.027$~eV $\approx k_BT$.
This value should not be read as a fixed material constant, however.
It is small on the scale of typical electronic-structure uncertainties,
and its absolute magnitude depends on the exchange--correlation
functional, the van der Waals treatment, and the interlayer stacking, as
quantified in Sec.~\ref{sec:robustness}. The hybrid HSE06 functional, for
example, places it near $0.1$~eV.
What matters for transport is not the precise number but the
\emph{proximity to the K--$\Gamma$ crossover}.
When $\Delta_{K\Gamma}$
is of order $k_BT$, both valleys are thermally accessible and
gate-induced Fermi filling redistributes carriers between them,
enabling a valley-driven mobility modulation that is absent in the
monolayer
($\Delta_{K\Gamma} \approx 0.5$~eV $\gg k_BT$) and that fades once the
splitting grows large.

\subsection{Valley Effective Masses}
\label{sec:effmass}

The hole effective masses at the $K$ and $\Gamma$ points, $m^*_K$ and
$m^*_\Gamma$, are extracted from parabolic fits to the DFT+SOC band
structure near each valence-band maximum.
The results at zero strain are summarized in Table~S4 of the
Supplemental Material~\cite{supplemental}.

The $K$-valley mass increases slightly with layer number
($m^*_K \approx 0.37$--$0.42\,m_0$), consistent with inter-layer
hybridization progressively modifying the spin-orbit-split upper band.
The $\Gamma$-valley mass varies strongly. It is largest in the
monolayer ($m^*_\Gamma \approx 3.37\,m_0$), and decreases toward
$1.00\,m_0$ (bilayer) and $0.78\,m_0$ (trilayer) as interlayer
coupling enhances the dispersion at $\Gamma$.
The mass hierarchy $m^*_\Gamma > m^*_K$ is corroborated by SdH
measurements in trilayer WSe$_2$, which yield
$m^*_K = 0.5\,m_0$ and $m^*_\Gamma = 1.2\,m_0$\cite{Movva2018_valleyPRL}.
The larger experimental $m^*_\Gamma$ relative to the present GGA-PBE value
($0.78\,m_0$) is consistent with the known tendency of GGA to
overestimate band dispersion at $\Gamma$.

The resulting effective-mass ratio $r_m = m^*_\Gamma / m^*_K$
decreases from $\approx 9.0$ (monolayer) to $\approx 2.5$ (bilayer)
and $\approx 1.9$ (trilayer). It sets the valley mobility ratio directly
and, together with the valley degeneracy introduced below, the valley
occupation.

Beyond the effective masses, the two valleys differ in their
\emph{degeneracy}. The bilayer is centrosymmetric (2H stacking), so
inversion together with time reversal renders every band spin-degenerate:
the $K$-valley maximum is a spin-degenerate doublet exhibiting hidden
spin--layer locking, with opposite spin polarization on the two layers,
as established for bulk and few-layer
WSe$_2$\cite{Riley2014_WSe2spinvalley,Jones2014_spinlayer,GongZ2013_magnetoelectric}.
Counting the two inequivalent valleys $K$ and $K'$, the $K$ manifold then
carries a total degeneracy $g_K=4$, whereas the single $\Gamma$ pocket
carries $g_\Gamma=2$. The state density entering the occupation is
$N_{\mathrm{2D},\nu}\propto g_\nu m^*_\nu$, so the quantity governing the
valley balance is the \emph{state-density} ratio
$r_N \equiv N_{2D,\Gamma}/N_{2D,K} = (g_\Gamma/g_K)\,r_m
= \tfrac{1}{2}\,r_m$ in the bilayer---half the value implied by the mass
ratio alone---whereas the mobility ratio
$\mu_\Gamma/\mu_K = m^*_K/m^*_\Gamma = 1/r_m$ is set by the \emph{mass}
ratio and is independent of the degeneracy.
The non-centrosymmetric
monolayer and trilayer are instead spin--valley locked, with $g_K=2$
(so $r_N=r_m$ there). The fourfold $K$ degeneracy is protected by the combined inversion and
time-reversal (PT) symmetry.
The same layer symmetry suppresses
the antisymmetric interlayer potential that would otherwise Stark-shift
the valley landscape, and a symmetry-breaking field lifts both
(Sec.~\ref{sec:discussion_efield}).

Both $m^*_K$ and $m^*_\Gamma$ decrease monotonically with tensile
biaxial strain (see Fig.~S2 in the Supplemental
Material~\cite{supplemental} and Tables~S1--S3 therein).
For device simulations at intermediate strain values, $\Delta_{K\Gamma}$
and $m^*$ are obtained by linear interpolation between the seven
DFT-calculated points. The smooth, monotonic variation of both
quantities with strain ensures that interpolation errors are negligible.
The SOC-derived splitting at the $K$ point is
467--482~meV across all layer
numbers\cite{Zhu2011_SOC,Riley2014_WSe2spinvalley,LeBoeuf2015_SOC}
and is essentially strain-independent at the level of the present
calculations. For the centrosymmetric bilayer this is the energy
separation between the upper and lower spin--layer doublets at $K$, not a
lifting of the PT degeneracy \emph{within} the occupied upper doublet
(which remains twofold; hence $g_K=4$). This lower manifold lies
${\approx}0.48$~eV below the valence-band edge and is thermally
inaccessible at room temperature, justifying its omission from the
two-pocket ($K$, $\Gamma$) transport model.

\section{Analytical Device Model}
\label{sec:model}

\begin{figure}[tbh]
  \centering
  \includegraphics[width=0.9\columnwidth]{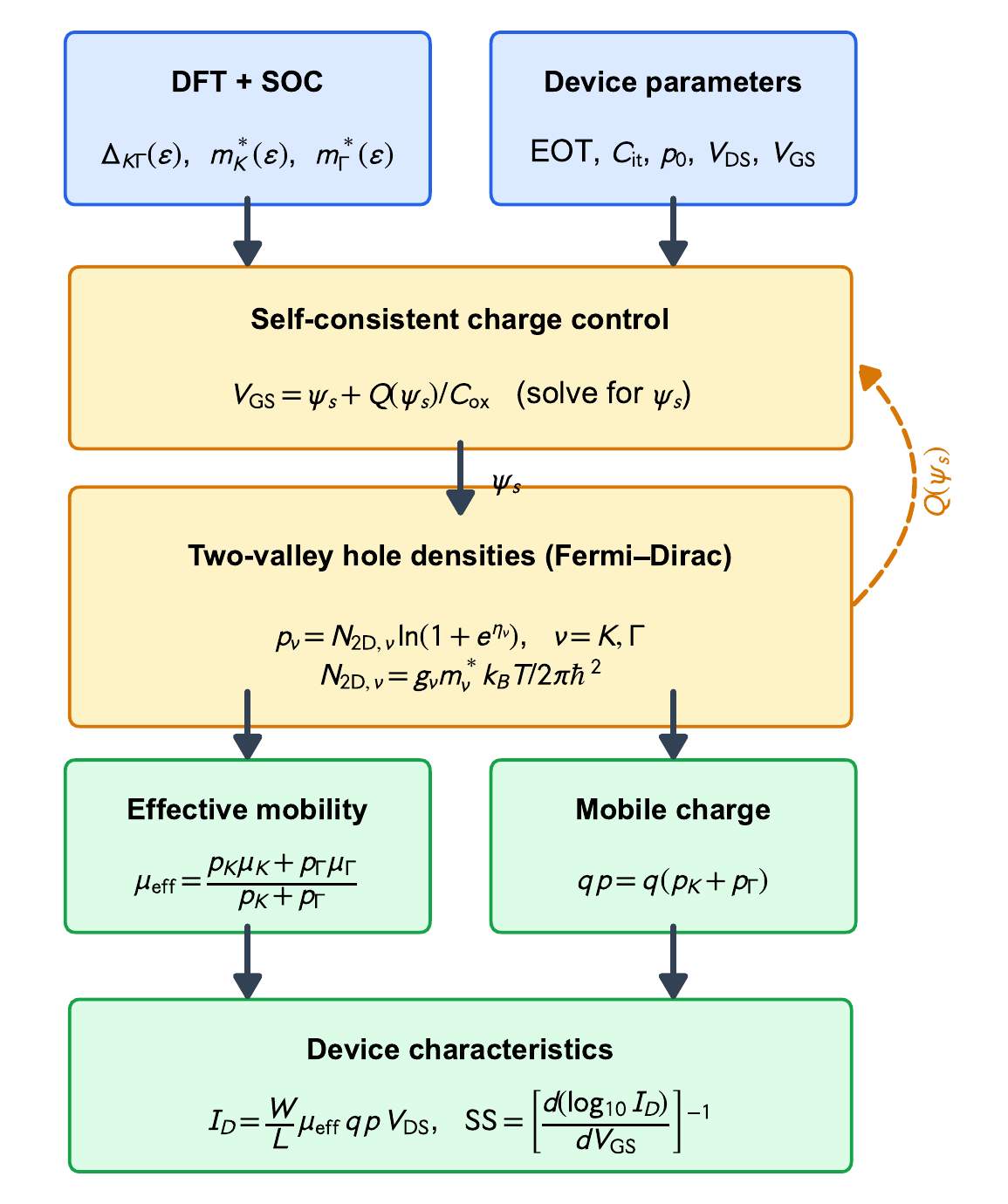}
  \caption{Flow diagram of the analytical two-valley transport model.
    Blue: inputs. Amber: the self-consistent core. Green: derived
    device characteristics.
    The DFT+SOC calculation supplies the valley splitting
    $\Delta_{K\Gamma}(\varepsilon)$ and the valley effective masses
    $m^*_K(\varepsilon)$, $m^*_\Gamma(\varepsilon)$, which together
    with the valley degeneracies $g_\nu$ (layer-dependent: $g_K=4$,
    $g_\Gamma=2$ for the centrosymmetric 2H bilayer shown here; $g_K=2$
    for the non-centrosymmetric monolayer and trilayer) and the device
    parameters enter the charge-control relation
    $V_\mathrm{GS}=\psi_s+Q(\psi_s)/C_\mathrm{ox}$
    [Eq.~(\ref{eq:vgs})].
    This relation is implicit, because $Q$ itself depends on $\psi_s$
    through the valley-resolved hole densities
    [Eqs.~(\ref{eq:pK}), (\ref{eq:pG})].
    The dashed loop indicates the self-consistent solution for
    $\psi_s$.
    The converged densities then determine the effective mobility
    $\mu_\mathrm{eff}$ [Eq.~(\ref{eq:mueff})] and the mobile charge
    $q\,p$, which together give the drain current
    [Eq.~(\ref{eq:Id})] and the subthreshold swing.}
  \label{fig:flow}
\end{figure}

The complete self-consistent computational workflow is illustrated in
Fig.~\ref{fig:flow}, and Table~\ref{tab:params} summarizes how each
quantity enters the model: which are taken from DFT, which are
externally assumed model parameters, and which are solved for.

\begin{table*}[tb]
  \caption{Classification of the quantities entering the two-valley
    device model. DFT inputs are computed in Sec.~\ref{sec:dft};
    model parameters are externally specified. Solved quantities
    follow from the self-consistent solution of
    Eq.~(\ref{eq:vgs}).
    EOT denotes the effective (total) GAA equivalent oxide thickness,
    so that $C_\mathrm{ox}$ is the total gate-to-channel capacitance per
    projected channel area (Eq.~\ref{eq:vgs}).}
  \label{tab:params}
  \begin{ruledtabular}
  \begin{tabular}{lll}
    Category & Quantity & Value / origin \\
    \colrule
    DFT input & $\Delta_{K\Gamma}(\varepsilon)$ &
      $0.0265$~eV (bilayer, $\varepsilon=0$) \\
     & $m^*_K(\varepsilon)$ & $0.395\,m_0$ (bilayer, $\varepsilon=0$) \\
     & $m^*_\Gamma(\varepsilon)$ & $1.00\,m_0$ (bilayer, $\varepsilon=0$) \\
     & $g_K,\ g_\Gamma$ & $4,\ 2$ (2H bilayer; Sec.~\ref{sec:effmass}) \\
     & $r_m = m^*_\Gamma/m^*_K$ & $2.53$ (effective-mass ratio) \\
     & $r_N = (g_\Gamma/g_K)\,r_m$ & $1.27$ (state-density ratio) \\
    \colrule
    Model & $\mu_K$ & $100$~cm$^2$V$^{-1}$s$^{-1}$ (experiment) \\
    parameter & $\mu_\Gamma$ & $\mu_K/r_m$
      [Eq.~(\ref{eq:muG})] \\
     & $p_0$ & $10^{11}$~cm$^{-2}$ (sensitivity in
       Ref.~\cite{supplemental}) \\
     & $C_\mathrm{it}/C_\mathrm{ox}$ & $0$--$1$ (parametric) \\
     & EOT & $0.7$~nm \\
     & $V_\mathrm{DS}$ & $50$~mV \\
     & $W/L$ & $1$ (square channel, $W=L$) \\
    \colrule
    Solved & $\psi_s(V_\mathrm{GS})$ & Eq.~(\ref{eq:vgs}),
      self-consistent \\
     & $p_K,\ p_\Gamma$ & Eqs.~(\ref{eq:pK}), (\ref{eq:pG}) \\
     & $\mu_\mathrm{eff}$ & Eq.~(\ref{eq:mueff}) \\
     & $I_D$ & Eq.~(\ref{eq:Id}) \\
     & SS & $[d(\log_{10}I_D)/dV_\mathrm{GS}]^{-1}$ \\
  \end{tabular}
  \end{ruledtabular}
\end{table*}

\subsection{Device Geometry and Electrostatics}

A GAA FET is considered in which a few-layer WSe$_2$ nanosheet serves
as the channel, fully surrounded by a gate dielectric with
EOT $= 0.7$~nm.
The electrostatic relationship between the gate voltage
$V_\mathrm{GS}$ and the surface potential $\psi_s$ is governed by
\begin{equation}
  V_\mathrm{GS} = \psi_s + \frac{Q(\psi_s)}{C_\mathrm{ox}},
  \label{eq:vgs}
\end{equation}
where $\varepsilon_0$ is the vacuum permittivity, $\varepsilon_\mathrm{ox}$
the relative permittivity of the gate dielectric, and
$C_\mathrm{ox} = \varepsilon_0 \varepsilon_\mathrm{ox} /
\mathrm{EOT}$ is the \emph{total} effective gate-to-channel capacitance
per projected channel area of the wrap-around geometry, so that the
stated EOT is an effective GAA equivalent oxide thickness rather than a
single-interface oxide thickness; $Q(\psi_s)$ is the total areal charge
density. (For a symmetric wrap-around gate the individual interface
capacitances add. The value $C_\mathrm{ox}=4.9\ \mu$F\,cm$^{-2}$ used
here corresponds, e.g., to two interfaces of ${\approx}1.4$~nm EOT each.
Doubling $C_\mathrm{ox}$ leaves the equilibrium valley occupation and the
thermionic-limit subthreshold swing unchanged and sharpens the
$V_\mathrm{GS}$ scale only modestly, without altering any conclusion.)
A positive gate-voltage convention for hole accumulation is adopted
here. This corresponds to negative gate bias in experimental p-type
FET operation.
The off-state current is evaluated at $V_\mathrm{GS} = 0$~V, where
the device is in the subthreshold regime.
Because $Q$ depends on $\psi_s$ through the valley-resolved carrier
densities derived below, Eq.~(\ref{eq:vgs}) is an implicit equation
for $\psi_s$ and is solved self-consistently at each $V_\mathrm{GS}$
by bracketed root finding (Brent's method, tolerance $10^{-10}$).
All reported quantities are evaluated from the converged $\psi_s$.

Two points about the scope of the model should be stated explicitly.
First, Eqs.~(\ref{eq:vgs}) and (\ref{eq:Id}) assume the
gradual-channel and long-channel limits at low drain bias
($V_\mathrm{DS} = 50$~mV), so that the electrostatics are effectively
one-dimensional and short-channel effects such as drain-induced
barrier lowering are absent.
The model is therefore intended to isolate valley-driven \emph{trends}
rather than to provide quantitative predictions for aggressively
scaled channel lengths.
An order-of-magnitude electrostatic estimate of the natural screening
length,
$\lambda \sim \sqrt{(\varepsilon_\mathrm{ch}/\varepsilon_\mathrm{ox})\,t_\mathrm{ch}\,\mathrm{EOT}} \approx 0.9$~nm
(with $\varepsilon_\mathrm{ch}$ the channel relative permittivity and
$t_\mathrm{ch}$ the channel thickness),
suggests that appreciable short-channel coupling is expected primarily at
channel lengths of a few nanometres or below. A self-consistent \emph{finite-bias}
extension, which confirms that the valley-driven trends persist as the
carrier density and valley occupation vary along the channel, is given
in the Supplemental Material~\cite{supplemental}.
Second, the valley-redistribution mechanism itself is
architecture-independent. It originates in the band structure and
applies equally to planar, FinFET, and nanosheet channels.
The GAA geometry is adopted here for three reasons.
It realizes the strong-electrostatic-control limit relevant to the
subthreshold regime, in which the electrostatic penalty
$(C_Q+C_\mathrm{it})/C_\mathrm{ox}$ is minimized.
Here $C_Q$ is the quantum capacitance and $C_\mathrm{it}$ the
interface-trap capacitance, both introduced below.
Full encapsulation by high-$\kappa$ dielectrics may help stabilize the
compressive strain identified below as favorable for maximizing the
on-state mobility.
Finally, its symmetric wrap-around gate suppresses the interlayer
Stark renormalization discussed in Sec.~\ref{sec:discussion_efield}.

\subsection{Two-Valley Hole Density}

The charge density is written as
\begin{equation}
  Q(\psi_s) = q\,p(\psi_s) + C_\mathrm{it}\,\psi_s,
  \label{eq:charge}
\end{equation}
where $q$ is the elementary charge, $C_\mathrm{it}$ is the
interface-trap capacitance, and $p(\psi_s)$ is the total hole
density.
$C_\mathrm{it}$ is not obtained from first principles. It is treated
as a parametric quantity and swept, so that the sensitivity of the
results to interface quality can be assessed independently of any
particular fabrication process.
In the standard small-signal picture it is related to the interface
trap density $D_\mathrm{it}$ by $C_\mathrm{it} = q^2 D_\mathrm{it}$ when
$D_\mathrm{it}$ is expressed per unit energy in joules. Equivalently,
$C_\mathrm{it} = q\,D_\mathrm{it}$ when $D_\mathrm{it}$ is quoted in the
conventional units of cm$^{-2}$eV$^{-1}$, which is the convention used
below.
With $C_\mathrm{ox} = 4.9\ \mu$F\,cm$^{-2}$ for EOT $=0.7$~nm, the
range $C_\mathrm{it}/C_\mathrm{ox} = 0$--$1$ explored here corresponds
to $D_\mathrm{it} = 0$--$3\times10^{13}$~cm$^{-2}$eV$^{-1}$.
Reported values for high-$\kappa$ dielectrics on 2D channels lie in
the $10^{12}$--$10^{13}$~cm$^{-2}$eV$^{-1}$
range,\cite{Illarionov2020_Ditreview,Knobloch2020_trapping,Ghatak2015_trapping}
i.e.\ $C_\mathrm{it}/C_\mathrm{ox} \approx 0.03$--$0.3$, so the
parameter range used here brackets realistic interface quality.
For a 2D parabolic band, the Fermi-Dirac integral has a closed form,
and the per-valley hole densities are
\begin{align}
  p_K(\psi_s) &= N_{2D,K}\,\ln\!\left(1 + e^{\eta_K(\psi_s)}\right),
    \label{eq:pK} \\
  p_\Gamma(\psi_s) &= N_{2D,\Gamma}\,\ln\!\left(1 +
    e^{\eta_K(\psi_s) - \Delta_{K\Gamma}/k_BT}\right), \label{eq:pG}
\end{align}
where
\begin{equation*}
  N_{2D,\nu} = \frac{g_\nu\,m^*_\nu\,k_BT}{2\pi\hbar^2}
\end{equation*}
is the 2D density-of-states prefactor for valley $\nu$ and $g_\nu$ is its
total degeneracy ($g_K=4$, $g_\Gamma=2$ for the 2H bilayer;
Sec.~\ref{sec:effmass}). The ratio that governs the valley balance is thus
the \emph{state-density} ratio
$r_N \equiv N_{2D,\Gamma}/N_{2D,K} = (g_\Gamma/g_K)\,r_m$, distinct from
the \emph{effective-mass} ratio $r_m \equiv m^*_\Gamma/m^*_K$. For the
bilayer $r_N = \tfrac{1}{2}\,r_m$.
Finally $\eta_K(\psi_s) = \eta_{K,0} + q\psi_s/k_BT$, where the single
equilibrium reduced Fermi level $\eta_{K,0}$ is fixed by requiring that
the \emph{total} equilibrium hole density reproduce the specified value
$p_0$, i.e.\ $p_K(0)+p_\Gamma(0)=p_0$ with one common Fermi level.
Thus $p_0$ is the total equilibrium hole density (set by doping or the
gate work function), and a single Fermi level populates both valleys.
The total hole density at bias is $p = p_K + p_\Gamma$.
In the non-degenerate limit $p_0 \ll N_{2D,K}$, Eqs.~(\ref{eq:pK}) and
(\ref{eq:pG}) reduce to the Boltzmann expressions
$p_K = p_{K,0}\,e^{q\psi_s/k_BT}$ and
$p_\Gamma = r_N\,p_{K,0}\,e^{(q\psi_s-\Delta_{K\Gamma})/k_BT}$,
where the equilibrium $K$-valley density
$p_{K,0} = p_0/(1+r_N\,e^{-\Delta_{K\Gamma}/k_BT})$ follows
from the total-density constraint.
Here $q\psi_s$ and $\Delta_{K\Gamma}$ are both energies, the former
expressed through the surface potential $\psi_s$ and the latter taken
directly from DFT in eV.
When $\Delta_{K\Gamma} \gg k_BT$ (monolayer limit), the $\Gamma$-valley
term is exponentially suppressed and single-valley $K$ transport is
recovered.
When $\Delta_{K\Gamma} \lesssim k_BT$ (bilayer at room temperature),
both valleys contribute from the outset.

\subsection{Quantum Capacitance and Subthreshold Swing}

The differential hole charge with respect to surface potential defines
the quantum capacitance.
With Fermi-Dirac statistics, differentiating Eqs.~(\ref{eq:pK})
and (\ref{eq:pG}) gives
\begin{equation}
  C_Q = q\,\frac{\partial p}{\partial \psi_s}
      = \frac{q^2}{k_BT}
        \sum_{\nu=K,\Gamma}
        N_{2D,\nu}\,f^{\mathrm{FD}}_\nu,
  \label{eq:CQ}
\end{equation}
where $f^{\mathrm{FD}}_\nu = (1+e^{-\eta_\nu})^{-1}$ is the Fermi--Dirac
occupation at the edge of valley $\nu$, with $\eta_K$ as defined above and
$\eta_\Gamma = \eta_K - \Delta_{K\Gamma}/k_BT$.
Equation~(\ref{eq:CQ}) follows directly from
$\partial\ln(1+e^{\eta_\nu})/\partial\eta_\nu = f^{\mathrm{FD}}_\nu$ together with
$\partial\eta_\nu/\partial\psi_s = q/k_BT$.
Equivalently, integrating $q^2 g_{2D,\nu}(-\partial f^{\mathrm{FD}}/\partial E)$
over energy gives the same expression, where
$g_{2D,\nu} = N_{2D,\nu}/k_BT = g_\nu m^*_\nu/(2\pi\hbar^2)$ is the
energy-independent two-dimensional density of states.
In the non-degenerate limit $f^{\mathrm{FD}}_\nu \to 0$,
Eq.~(\ref{eq:CQ}) reduces to $C_Q = q^2 p/(k_BT)$, whereas in the
degenerate limit $f^{\mathrm{FD}}_\nu \to 1$ it saturates at the density-of-states
value $q^2\sum_\nu N_{2D,\nu}/k_BT$, as expected for a
two-dimensional band.
In the non-degenerate subthreshold regime, where the valley fraction
and hence $\mu_\mathrm{eff}$ are nearly gate-independent so that
$I_D \propto p$, the subthreshold swing is
\begin{equation}
  \mathrm{SS} = \frac{dV_\mathrm{GS}}{d(\log_{10} I_D)}
  = \ln(10)\frac{k_BT}{q}
    \left(1 + \frac{C_Q + C_\mathrm{it}}{C_\mathrm{ox}}\right).
  \label{eq:SS}
\end{equation}
Deep in the subthreshold regime, $C_Q \ll C_\mathrm{ox}$, so
$\mathrm{SS} \to \ln(10)\,k_BT/q \approx 60$~mV~dec$^{-1}$,
regardless of layer number.
The layer-number independence therefore has a simple
origin\cite{Luryi1988_QC,Bennett2023_QC}. A larger valley density of
states increases $C_Q$, but $C_Q$ enters Eq.~(\ref{eq:SS}) only through
the ratio $C_Q/C_\mathrm{ox}$. As long as $C_Q \ll C_\mathrm{ox}$ this
ratio is negligible and the swing is unchanged.
It should be emphasised that $C_Q$ can only degrade the swing, never
improve it, so the near-ideal value obtained here reflects the
smallness of
$C_Q/C_\mathrm{ox}$ rather than any active screening mechanism.
Equation~(\ref{eq:SS}) is used here for physical interpretation only.
In all numerical results reported below, SS is extracted directly from
the computed $I_D$--$V_\mathrm{GS}$ characteristics as
$[d(\log_{10} I_D)/dV_\mathrm{GS}]^{-1}$, so that no approximation to
$C_Q$ enters the reported device characteristics.

\subsection{Drain Current and Effective Mobility}

The drain current in the linear regime is modeled as
\begin{equation}
  I_D = \frac{W}{L}\,\mu_\mathrm{eff}\,q\,p(\psi_s)\,V_\mathrm{DS},
  \label{eq:Id}
\end{equation}
where $W/L$ is the channel aspect ratio.
Only the \emph{mobile} charge $q\,p$ enters the current.
The trapped charge $C_\mathrm{it}\psi_s$ appearing in
Eq.~(\ref{eq:charge}) contributes to the electrostatics through the
charge-control relation, Eq.~(\ref{eq:vgs}), and therefore degrades the
subthreshold swing, but it is immobile and carries no current.
All currents reported in this work are evaluated for $W/L = 1$, i.e.\
for a square channel with $W = L$. This choice is a current
\emph{normalization} only and is logically separate from the
long-channel \emph{assumption} ($L$ much larger than the electrostatic
scaling length) under which Eqs.~(\ref{eq:vgs}) and (\ref{eq:Id}) are
formulated.
Currents for any other geometry follow by multiplying by the desired
$W/L$.
Because every conclusion drawn here concerns either a ratio
($I_\mathrm{on}/I_\mathrm{off}$) or a relative trend with strain or
layer number, none of them depends on this normalization.
The density-weighted effective mobility is
\begin{equation}
  \mu_\mathrm{eff} = \frac{p_K\,\mu_K + p_\Gamma\,\mu_\Gamma}
                         {p_K + p_\Gamma}.
  \label{eq:mueff}
\end{equation}
Here $\mu_K$ and $\mu_\Gamma$ are the intravalley mobilities for the
$K$ and $\Gamma$ valleys, respectively.
$\mu_K = 100$~cm$^2$\,V$^{-1}$\,s$^{-1}$ is adopted, consistent with
experimental reports for monolayer
WSe$_2$,\cite{Allain2014_WSe2mobility,Pradhan2015_WSe2Hall,Gunst2025_WSe2mobility}
and $\mu_\Gamma$ is estimated by assuming a common intravalley
scattering time $\tau$ so that $\mu \propto 1/m^*$:
\begin{equation}
  \mu_\Gamma = \mu_K \frac{m^*_K}{m^*_\Gamma} = \frac{\mu_K}{r_m}.
  \label{eq:muG}
\end{equation}
Using the DFT effective masses (Table~S4 in the Supplemental
Material~\cite{supplemental}), this gives
$\mu_\Gamma \approx 11$~cm$^2$\,V$^{-1}$\,s$^{-1}$
(monolayer, $r_m=9.01$),
$\approx 40$~cm$^2$\,V$^{-1}$\,s$^{-1}$ (bilayer, $r_m=2.53$),
and $\approx 53$~cm$^2$\,V$^{-1}$\,s$^{-1}$ (trilayer,
$r_m=1.89$).
Under the equal-scattering-time approximation,
$m^*_\Gamma > m^*_K$\cite{Wickramaratne2014_indirect} gives
$\mu_\Gamma < \mu_K$, and $\mu_\mathrm{eff}$ decreases
monotonically as the $\Gamma$-valley fraction increases.

\section{Results}
\label{sec:results}

\subsection{Effective Mobility: Layer-Number and Valley-Population Tuning}

The primary goal of this work is to identify the \emph{relative trends}
driven by $\Delta_{K\Gamma}$---specifically how layer number and strain
shift the valley balance---rather than to predict absolute mobility
values.
The equal-scattering-time approximation [Eq.~(\ref{eq:muG})] is
sufficient for this purpose, and its limitations are discussed in
Sec.~\ref{sec:discussion}.

\begin{figure}[tb]
  \includegraphics[width=0.95\columnwidth]{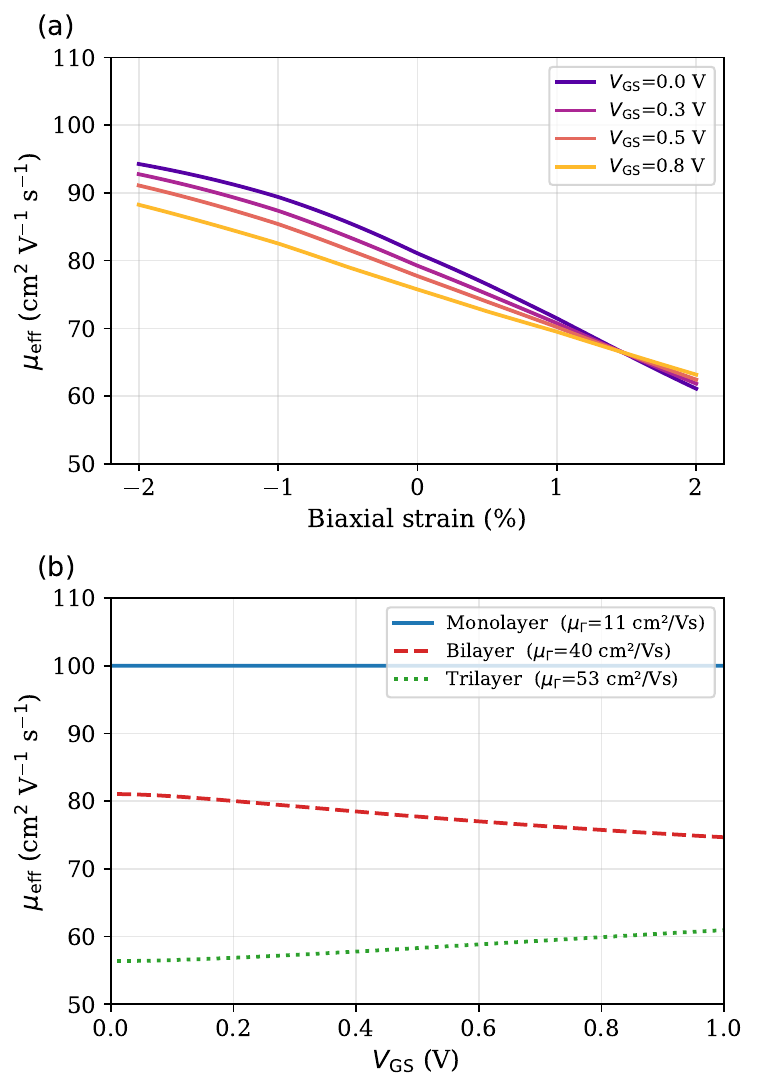}
  \caption{Effective hole mobility $\mu_\mathrm{eff}$
    of WSe$_2$ GAA FETs.
    (a)~$\mu_\mathrm{eff}$ of the bilayer as a function of biaxial
    strain at fixed gate voltages $V_\mathrm{GS} = 0.0, 0.3, 0.5,
    0.8$~V.
    Compressive (tensile) strain suppresses (enhances) $\Gamma$-valley
    occupation, providing a continuous and reversible means to tune mobility
    without degrading the subthreshold swing.
    (b)~$\mu_\mathrm{eff}$ as a function of $V_\mathrm{GS}$ for
    monolayer (solid), bilayer (dashed), and trilayer (dotted) at zero
    strain.
    The monotonic decrease with layer number reflects the increasing
    $\Gamma$-valley occupation (see Fig.~S3(a), Supplemental
    Material~\cite{supplemental}).
    Absolute values depend on the assumed $\mu_K$ and the
    equal-scattering-time approximation. The physically significant
    result is the monotonic trend.}
  \label{fig:mueff}
\end{figure}

Figure~\ref{fig:mueff}(a) shows $\mu_\mathrm{eff}$ of the bilayer
as a function of biaxial strain at several fixed gate voltages.
The effective mobility is governed by the valley population.
By tuning $\Delta_{K\Gamma}$ via strain, holes redistribute between
the light $K$ valley and the heavy $\Gamma$ valley, directly
controlling $\mu_\mathrm{eff}$.
Compressive strain shifts the population toward $K$, recovering
$K$-valley-like mobility, while tensile strain drives the K--$\Gamma$
crossover and lowers $\mu_\mathrm{eff}$.
The strain sensitivity is largest at \emph{low} $V_\mathrm{GS}$: the
span of $\mu_\mathrm{eff}$ between $-2\%$ and $+2\%$ strain falls from
$33$~cm$^2$\,V$^{-1}$\,s$^{-1}$ at $V_\mathrm{GS}=0$ to
$25$~cm$^2$\,V$^{-1}$\,s$^{-1}$ at $V_\mathrm{GS}=0.8$~V.
In the non-degenerate regime the valley ratio is
$p_\Gamma/p_K = r_N\,e^{-\Delta_{K\Gamma}/k_BT}
= \tfrac{1}{2}\,r_m\,e^{-\Delta_{K\Gamma}/k_BT}$ for the bilayer
and is therefore exponentially sensitive to strain, whereas at higher
carrier density Fermi-Dirac filling of both valleys progressively
compresses that response.

The four curves intersect at $\varepsilon \approx +1.5\%$, the
K--$\Gamma$ crossover. There $\Delta_{K\Gamma}=0$, the valley ratio
reduces to $r_N = \tfrac{1}{2}\,r_m$ irrespective of $\psi_s$, and
$\mu_\mathrm{eff}$ becomes gate-independent.
This gate-independence of $\mu_\mathrm{eff}$ at the crossover reflects a
broader separation of mobility engineering from the subthreshold
behaviour in the bilayer platform, developed in
Sec.~\ref{sec:discussion}\cite{Desai2014_WSe2strain,Lau2024_straintronics}.
The underlying design principle is concise:
\emph{valley-engineering sensitivity is largest in the small-splitting
regime}, peaking (in the Boltzmann limit) near
$\Delta_{K\Gamma}=k_BT\ln r_N$
($\approx 6$~meV for the bilayer).
This regime can be reached through
strain, stacking, and dielectric engineering (Sec.~\ref{sec:robustness}).

Figure~\ref{fig:mueff}(b) shows $\mu_\mathrm{eff}$ as a function of
$V_\mathrm{GS}$ for monolayer, bilayer, and trilayer WSe$_2$ GAA FETs
at zero strain.
In the monolayer, the $\Gamma$ valley is fully suppressed
($\Delta_{K\Gamma} = 0.5031$~eV $\gg k_BT$), so holes reside entirely
in the light $K$ valley and
$\mu_\mathrm{eff} \approx \mu_K = 100$~cm$^2$\,V$^{-1}$\,s$^{-1}$
throughout.
In the bilayer, $\Delta_{K\Gamma} \approx k_BT$ gives substantial thermal
population of \emph{both} valleys---with the higher-degeneracy $K$
manifold ($g_K=4$; Sec.~\ref{sec:effmass}) in the majority---and a
$\Gamma$-valley occupation fraction
$f_\Gamma \equiv p_\Gamma/(p_K+p_\Gamma)$ rising from $31\%$ to
$42\%$ over $V_\mathrm{GS} = 0$--$1$~V, so that
$\mu_\mathrm{eff}$ falls from $81$ to
$75$~cm$^2$\,V$^{-1}$\,s$^{-1}$.
The trilayer is $\Gamma$-valley dominated from the outset
($f_\Gamma \approx 93\%$ at $V_\mathrm{GS}=0$).
Its gate dependence runs \emph{opposite} to that of the bilayer.
Because the $\Gamma$ valley is already the majority carrier reservoir,
raising the density brings the light $K$ valley into play, so
$f_\Gamma$ falls to $83\%$ and $\mu_\mathrm{eff}$ rises from $56$ to
$61$~cm$^2$\,V$^{-1}$\,s$^{-1}$ over the same range.
The monotonic ordering with layer number ($\mu_\mathrm{eff}$: monolayer
$>$ bilayer $>$ trilayer) therefore holds across the bias window shown,
the bilayer remaining above the trilayer throughout.
The $\Gamma$-valley occupation fraction for all layer numbers is shown
in Fig.~S3(a) of the Supplemental Material~\cite{supplemental}.

\subsection{$I_D$--$V_\mathrm{GS}$ Characteristics}

\begin{figure*}[tb]
  \includegraphics[width=\textwidth]{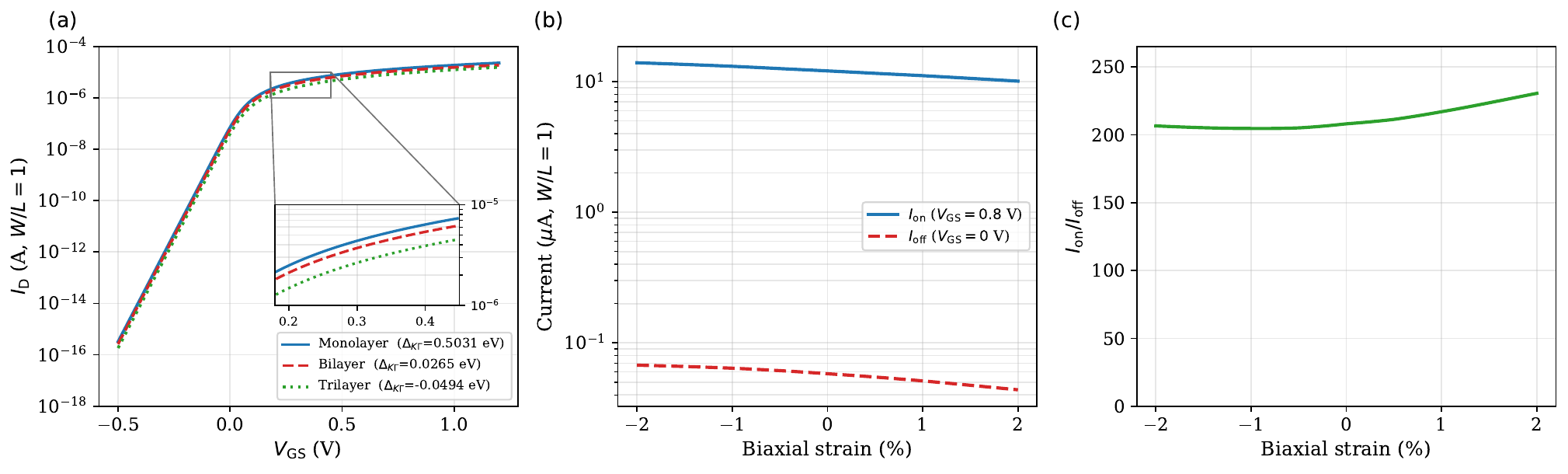}
  \caption{Calculated device characteristics of WSe$_2$ GAA FETs.
    (a)~$I_D$--$V_\mathrm{GS}$ characteristics (semilogarithmic scale)
    for monolayer, bilayer, and trilayer at zero strain.
    Device parameters: EOT $= 0.7$~nm, $p_0 = 10^{11}$~cm$^{-2}$,
    $V_\mathrm{DS} = 50$~mV, $C_\mathrm{it} = 0$, $W/L = 1$.
    (b)~On-current $I_\mathrm{on}$ ($V_\mathrm{GS} = 0.8$~V, blue
    solid) and off-current $I_\mathrm{off}$ ($V_\mathrm{GS} = 0$~V, red
    dashed) of bilayer WSe$_2$ as functions of biaxial strain, plotted
    on a single logarithmic axis.
    (c)~Corresponding on/off ratio $I_\mathrm{on}/I_\mathrm{off}$.
    Device parameters as in (a).
    Both $I_\mathrm{on}$ and $I_\mathrm{off}$ decrease with tensile
    strain, tracking the reduced $\mu_\mathrm{eff}$ at fixed total
    equilibrium density $p_0$; barrier modulation and contact-limited
    contributions are not included.
    The on/off ratio is therefore only weakly strain-dependent
    (${\approx}206$ at $-2\%$ to ${\approx}230$ at $+2\%$ for
    $p_0 = 10^{11}$~cm$^{-2}$)---strain acts on the mobility,
    not on the switching ratio.}
  \label{fig:id_combined}
\end{figure*}

Figure~\ref{fig:id_combined}(a) shows the calculated
$I_D$--$V_\mathrm{GS}$ characteristics for monolayer, bilayer, and
trilayer WSe$_2$ GAA FETs at zero strain on a semilogarithmic scale.
All three devices exhibit well-defined off-states and sharp turn-on
behavior.
The on-current at $V_\mathrm{GS} = 0.5$~V decreases from the monolayer
to the trilayer, consistent with the layer-number dependence of
$\mu_\mathrm{eff}$ shown in Fig.~\ref{fig:mueff}(b).
The minimum subthreshold slope is nearly identical for all layer numbers,
a consequence of $C_Q \ll C_\mathrm{ox}$ deep in the subthreshold regime,
as discussed in Sec.~\ref{sec:model} (see also Fig.~S3(b), Supplemental
Material~\cite{supplemental}).

Figure~\ref{fig:id_combined}(b) shows the on-current $I_\mathrm{on}$
($V_\mathrm{GS} = 0.8$~V) and the off-current $I_\mathrm{off}$
($V_\mathrm{GS} = 0$~V) of the bilayer as functions of biaxial strain,
and Fig.~\ref{fig:id_combined}(c) shows the resulting on/off ratio
$I_\mathrm{on}/I_\mathrm{off}$.
The valley population, tuned via strain, sets the effective mobility
and hence the on-state current. Tensile strain drives holes into the
heavy $\Gamma$ valley, lowering $\mu_\mathrm{eff}$ and reducing
$I_\mathrm{on}$, while compressive strain shifts the population toward
the light $K$ valley and raises $I_\mathrm{on}$.
The equilibrium hole density $p_0$ here is the \emph{total} density
$p_K(0)+p_\Gamma(0)$, fixed by a single equilibrium Fermi level. The
off-state current at $V_\mathrm{GS}=0$ is then
$I_\mathrm{off}\propto\mu_\mathrm{eff}(0)\,p_0$, so it is governed by the
same valley-averaged mobility as $I_\mathrm{on}$.
Consequently $I_\mathrm{off}$ follows the \emph{same} trend as
$I_\mathrm{on}$---both decrease under tensile strain---rather than moving
in opposite directions.
(The $I_\mathrm{off}$ reported here derives solely from this equilibrium
mobile charge; drain-induced barrier lowering, contact resistance, and
band-to-band tunneling are absent from the present model.)

At the on-state ($V_\mathrm{GS}=0.8$~V) the accumulated mobile density is
fixed largely by the gate electrostatics and is nearly
strain-independent---it varies by less than $1\%$ across
$\varepsilon=-2\%$ to $+2\%$, whereas $\mu_\mathrm{eff}$ varies by
${\approx}33\%$---so the strain modulation of
$I_\mathrm{on}=(W/L)\,\mu_\mathrm{eff}\,q\,p\,V_\mathrm{DS}$ is carried
almost entirely by $\mu_\mathrm{eff}$ rather than by the carrier number
$p$ [Fig.~S4(c), Supplemental Material~\cite{supplemental}].
Because $I_\mathrm{on}$ and $I_\mathrm{off}$ scale together with
$\mu_\mathrm{eff}$, the static on/off ratio is only weakly
strain-dependent. It varies from ${\approx}206$ at $-2\%$ to
${\approx}230$ at $+2\%$ [Fig.~\ref{fig:id_combined}(c)]---about $12\%$,
with a slight increase toward tensile strain.
This weak dependence is robust to the model inputs: varying $p_0$ over
$10^{10}$--$10^{12}$~cm$^{-2}$ rescales the absolute currents but leaves
the on/off ratio at $-2\%$ within a factor $0.89$--$0.90$ of its $+2\%$
value, and varying the valley mobility ratio $\mu_\Gamma/\mu_K$ over
$0.2$--$0.8$ keeps that factor in the range $0.86$--$0.96$
(Supplemental Material~\cite{supplemental}).
The device-relevant result is therefore not a strain-driven improvement
of the on/off ratio but the electronic, strain- and gate-tunable control
of the effective mobility (and of $I_\mathrm{on}$), which is decoupled
from the subthreshold swing (Sec.~\ref{sec:results}~A;
Fig.~S4~\cite{supplemental}). The on/off ratio quoted here is a model
comparison metric---set by the assumed $p_0$ and the $V_\mathrm{GS}=0$
reference---rather than a prediction of a specific device's switching
ratio.
The underlying $I_D$--$V_\mathrm{GS}$ characteristics for each strain
value are provided in the Supplemental Material (Fig.~S4(a)~\cite{supplemental}).

\subsection{Comparison with Experiment}

Since no experimental $I_D$--$V_\mathrm{GS}$ data for bilayer WSe$_2$
GAA FETs are currently available, the electrostatic framework of the
model---specifically its subthreshold slope---is checked for consistency
against monolayer WSe$_2$ data.
The purpose and the limits of this comparison should be stated
explicitly.
What is tested is the \emph{charge-control and electrostatic} part of
the model---Eqs.~(\ref{eq:vgs})--(\ref{eq:SS})---which depends on
$C_\mathrm{ox}$, $C_Q$, and $C_\mathrm{it}$ but not on the valley
structure, and which therefore carries over unchanged from the
monolayer to the bilayer.
The monolayer is the cleanest available test of this part,
because its large splitting ($\Delta_{K\Gamma}\approx0.5$~eV) places
it in the single-valley limit, so that any discrepancy is attributable
to electrostatics rather than to valley physics.
What is \emph{not} tested by Fig.~\ref{fig:validation}(a) is the two-valley
transport prediction itself, which is specific to the bilayer.
Recent measurements on bilayer WSe$_2$ p-FETs\cite{Ghosh2025_bilayerWSe2FET}
and \textit{ab initio} quantum-transport calculations for
mono- and bilayer WSe$_2$ channels\cite{Wang2025_WSe2NEGF} provide
independent support for the device platform, but do not yet resolve
the valley-resolved mobility.
The bilayer predictions of Sec.~\ref{sec:results} therefore constitute
testable theoretical proposals awaiting dedicated experiments in which
strain and gate bias are varied simultaneously.

Figure~\ref{fig:validation}(a) compares the model prediction with the
experimental $I_D$--$V_\mathrm{GS}$ data of
Ref.~\cite{Fang2012_WSe2pFET} for a monolayer WSe$_2$ p-FET with a
ZrO$_2$ top gate (EOT $= 5.5$~nm, $L = 9.4$~$\mu$m,
$V_\mathrm{DS} = -0.05$~V).
Using the DFT-computed $\Delta_{K\Gamma} = 0.5031$~eV as the sole
first-principles input and setting
$\mu_K = 250$~cm$^2$\,V$^{-1}$\,s$^{-1}$ (taken from the reported
peak mobility\cite{Fang2012_WSe2pFET}) and
$C_\mathrm{it}/C_\mathrm{ox} = 0$, the model predicts an ideal SS of
60~mV~dec$^{-1}$, matching the near-ideal swing reported in that work.
The comparison is deliberately restricted to this slope.
The absolute current is not predicted, because it depends on the
unknown equilibrium hole density and, in the experiment, on a large
Schottky-barrier contact resistance in the Pd-contacted
devices\cite{Chuang2014_WSe2graphene,Tosun2014_WSe2CMOS} that the ideal
model omits.
The model curve is therefore aligned to the data by a rigid shift along
the gate-voltage axis (the flat-band voltage $V_\mathrm{FB}$) and a
single constant vertical scaling of the current, neither of which
affects the subthreshold slope being tested.
The dashed curve for $C_\mathrm{it}/C_\mathrm{ox} = 0.5$
(SS $\approx 90$~mV~dec$^{-1}$) illustrates how moderate
interface-trap density degrades the subthreshold slope, providing a
bound on the interface quality consistent with the experimental data.

\subsection{Effect of Interface Traps}

\begin{figure*}[tb]
  \includegraphics[width=\textwidth]{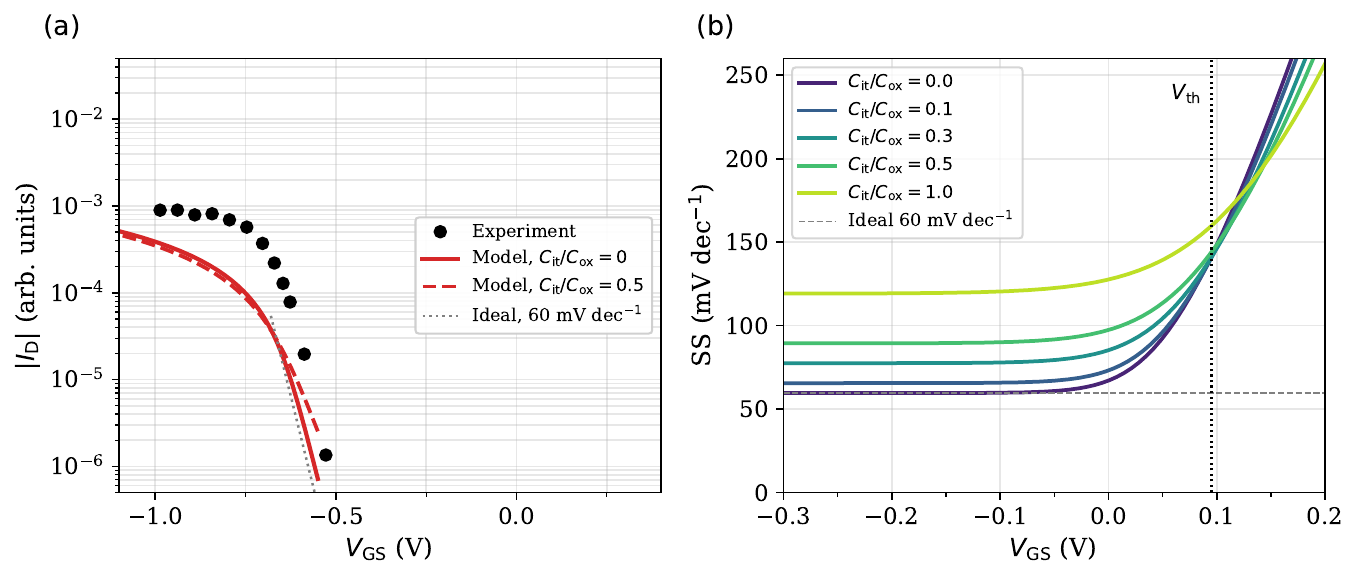}
  \caption{Consistency check of the electrostatic framework against the
    measured subthreshold slope, and sensitivity to
    interface traps.
    (a)~Comparison of the calculated subthreshold characteristics with
    experimental data for a monolayer WSe$_2$ p-FET
    [Ref.~\cite{Fang2012_WSe2pFET}, Fig.~4b therein].
    The comparison tests the subthreshold slope only. It does not test the
    two-valley transport prediction, which is specific to the bilayer
    (Sec.~\ref{sec:results}).
    The DFT value $\Delta_{K\Gamma} = 0.5031$~eV and
    $\mu_K = 250$~cm$^2$\,V$^{-1}$\,s$^{-1}$ are fixed; the model curve is
    aligned to the data by a rigid gate-voltage shift (flat-band voltage
    $V_\mathrm{FB}$) and a single constant vertical scaling of the current,
    neither of which affects the slope, so the current axis is in arbitrary
    units.
    The dotted line marks the ideal SS $= 60$~mV~dec$^{-1}$.
    (b)~Subthreshold swing as a function of gate voltage for the monolayer with
    $C_\mathrm{it}/C_\mathrm{ox} = 0, 0.1, 0.3, 0.5, 1.0$, restricted to the
    subthreshold region $V_\mathrm{GS} < V_\mathrm{th}$.
    The horizontal dashed line marks the thermionic limit of
    $\approx 60$~mV~dec$^{-1}$.}
  \label{fig:validation}
\end{figure*}

Figure~\ref{fig:validation}(b) shows the subthreshold swing SS as a function of
$V_\mathrm{GS}$ for the monolayer device with $C_\mathrm{it}/C_\mathrm{ox}$
from 0 to 1.0.
Increasing interface-trap density degrades SS from the ideal
$60$~mV~dec$^{-1}$ toward $120$~mV~dec$^{-1}$, in accordance with
Eq.~(\ref{eq:SS}).
The threshold voltage $V_\mathrm{th}$ shifts positively (toward larger
$V_\mathrm{GS}$) as interface traps act as an additional capacitive load.
The dependence of the minimum SS on $C_\mathrm{it}/C_\mathrm{ox}$
follows the analytical formula
SS $= \mathrm{SS}_0(1 + C_\mathrm{it}/C_\mathrm{ox})$, where
$\mathrm{SS}_0 \approx 60$~mV~dec$^{-1}$ is the trap-free thermionic
limit, and is shown
in the Supplemental Material (Fig.~S5(c)~\cite{supplemental}).
These results highlight that interface quality, rather than valley
structure, is the dominant extrinsic factor controlling SS in
WSe$_2$ GAA FETs, consistent with experimental
observations\cite{Ali2024_WSe2SS,Jayachandran2023_MoS2SS,Illarionov2020_Ditreview,Knobloch2020_trapping,Ghatak2015_trapping}.
The corresponding $I_D$--$V_\mathrm{GS}$ characteristics are provided
in the Supplemental Material (Fig.~S5(a)~\cite{supplemental}).

\section{Discussion}
\label{sec:discussion}

\subsection{Design Implications}

The key advance of this work is a design principle that
\emph{decouples} mobility modulation from the electrostatic factors that
govern the subthreshold swing.
Valley-population redistribution in bilayer WSe$_2$ controls the
effective mobility by redistributing holes among valleys of different
transport mobility, not through the carrier number or the gate
efficiency.
It therefore tunes $\mu_\mathrm{eff}$ and the on-current while leaving the
subthreshold swing at the thermionic limit.
This differs from strategies that tune mobility primarily by modifying
the scattering rate.
Here the scattering time of each valley is fixed, and the effective
mobility changes only because holes shift between the $K$ and $\Gamma$
valleys, which have different mobilities.

The present analysis leads to concrete design guidelines for WSe$_2$ GAA
FETs, organized by layer number.
A monolayer channel confines transport to the $K$ valley and therefore
provides the highest effective mobility, making it the natural choice
when maximum on-current is the priority.
A bilayer channel instead places the splitting in the
small-$\Delta_{K\Gamma}$ range where valley engineering is active.
There, compressive strain ($\varepsilon < 0$) suppresses the
$\Gamma$-valley occupation, recovers $K$-valley-like mobility, and raises
$I_\mathrm{on}$.
Because the off-state current scales with the same $\mu_\mathrm{eff}$ at
fixed equilibrium density, the on/off ratio stays nearly constant with
strain (Sec.~\ref{sec:results}~C), while the minimum subthreshold swing
remains near $60$~mV~dec$^{-1}$ well below threshold.
Strain thus provides electronic control of the mobility that is
decoupled from the switching slope, which is set instead by the
electrostatics and the interface quality.
A trilayer channel, being $\Gamma$-valley dominated, offers a lower
effective mobility in the present model.
For all three layer numbers the subthreshold swing stays near the
thermionic limit ($C_Q \ll C_\mathrm{ox}$), so the interface-trap
density, rather than the valley structure, is the primary extrinsic
factor that degrades it.

The valley-redistribution mechanism identified here is complementary
to the carrier-distribution engineering approach of
Ref.~\cite{Liang2023CPL}, which redistributes carriers spatially
through dual-gate electrostatics.
The present work instead controls the valley occupation ratio at the
band-structure level, offering an independent and gate-accessible
means to control mobility.

From a practical standpoint, biaxial strains in the range
$-2\%$ to $+2\%$ are within experimental reach in encapsulated
2D-material devices. Thermal expansion mismatch between WSe$_2$
and common gate dielectrics (HfO$_2$, Al$_2$O$_3$) during
processing typically introduces compressive strains of
$0.5$--$2\%$, and intentional substrate engineering or
flexible-support bending can extend this
range\cite{He2013,Conley2013,Schmidt2024_straintransfer,Carrascoso2021_strainTMD}.
In terms of device metrics, the model predicts that compressive strain
in the bilayer raises the on-current through $\mu_\mathrm{eff}$, while
the minimum subthreshold swing stays at ${\approx}60$~mV~dec$^{-1}$ at
every strain [Fig.~S4(b), Supplemental Material~\cite{supplemental}].
The distinctive signature of the valley-redistribution mechanism is
therefore this decoupling---$\mu_\mathrm{eff}$ (and $I_\mathrm{on}$)
tuned electronically by strain and gate bias while the switching slope
is left unchanged---rather than a change in the on/off ratio, which the
mechanism leaves nearly strain-independent.
Recent advances in WSe$_2$ device fabrication---including h-BN
encapsulation, high-$\kappa$ dielectric
integration,\cite{Smets2023_WSe2improvements,Chiu2025VLSI} and
scalable nanosheet
processing\cite{Loubet2017_nanosheet,Pan2025_TMDscaling}---place
direct experimental tests of these predictions within reach of
current technology.

\subsection{Experimental Signatures in Bilayer WSe$_2$}

The two-valley transport picture has direct experimental support.
SdH measurements in trilayer WSe$_2$ have confirmed simultaneous
K- and $\Gamma$-valley populations and shown that an applied
transverse electric field transfers holes from the $\Gamma$ valley to
the $K$ valley\cite{Movva2018_valleyPRL}.
Although these measurements were performed at cryogenic temperatures
in the trilayer (where the $\Gamma$ valley is the VBM throughout),
they establish that the K--$\Gamma$ redistribution mechanism is
experimentally accessible and valley populations respond to
electrostatic control.
Direct bilayer-specific FET measurements with simultaneous strain
control at room temperature are not yet available. The present model
makes several experimentally testable predictions.

First, the gate-voltage dependence of $\mu_\mathrm{eff}$ should differ
qualitatively between layer numbers, and the \emph{sign} of that
dependence is the sharpest discriminator.
The monolayer maintains a nearly constant $\mu_\mathrm{eff}$ across
the on-state, being $K$-valley dominated throughout.
The bilayer should show a monotonic \emph{decrease} with
$V_\mathrm{GS}$ as the heavier $\Gamma$ valley fills.
The trilayer should show a monotonic \emph{increase}, because it
starts $\Gamma$-dominated and the light $K$ valley is recruited as the
density rises.
A reproducible slope reversal between bilayer and trilayer would
constitute a characteristic signature consistent with the predicted
valley-redistribution mechanism, particularly if accompanied by the
expected strain dependence.
Second, biaxial strain should tune the on-current through the effective
mobility---compressive strain raising $\mu_\mathrm{eff}$ and hence
$I_\mathrm{on}$---while the subthreshold swing stays at the thermionic
limit. Because the off-state current at fixed equilibrium density scales
with the same $\mu_\mathrm{eff}$, it moves in the \emph{same} direction
as $I_\mathrm{on}$, so the on/off ratio is nearly strain-independent.
The feature that distinguishes valley redistribution from conventional
single-valley transport is therefore that $\mu_\mathrm{eff}$ (and
$I_\mathrm{on}$) is tunable by strain and gate bias while the switching
slope is unaffected---a strain-tunable mobility decoupled from the
electrostatic swing, rather than a change in the on/off ratio.

\subsection{Gate Symmetry: Carrier-Density Control versus Band
Modification}
\label{sec:discussion_efield}

The gate can influence the valley balance through two physically distinct
mechanisms, which must be clearly separated. First, at \emph{fixed}
$\Delta_{K\Gamma}$ the gate controls the valley \emph{occupation} through
the common chemical potential: raising the hole density fills the heavier
$\Gamma$ valley faster than the lighter $K$ valley, shifting the
population from $K$ toward $\Gamma$. This Fermi-filling mechanism requires
no change in the band structure; it is the physical content of the
two-valley model and the origin of the $V_\mathrm{GS}$ dependence of
$\mu_\mathrm{eff}$ in Fig.~\ref{fig:mueff}(b) and of every result in
Sec.~\ref{sec:results}. Second, a symmetry-breaking out-of-plane field
can modify the band structure itself through the interlayer Stark
effect\cite{Ramasubramaniam2011,Gong2013,Ross2013,Zheng2019_WSe2efield,Zhang2025NanoLett,Wei2025_twistedWSe2KG},
changing $\Delta_{K\Gamma}$; the results of Sec.~\ref{sec:results} rest
entirely on the first.

\begin{table*}[t]
  \caption{Comparison of gate-induced electrostatics in single-gate
    planar, dual-gate, and symmetric gate-all-around (GAA) bilayer
    WSe$_2$ transistors. The common-mode potential controls the total
    carrier density, whereas the antisymmetric (differential) component
    controls the interlayer displacement field and hence the interlayer
    Stark renormalization of $\Delta_{K\Gamma}$. A dual-gate device can
    select between these two actions through the sum and difference of
    its two gate voltages (for comparable top- and bottom-gate
    capacitances); the ideal symmetric GAA structurally realizes the
    common-mode limit.}
  \label{tab:gaa_planar}
  \begin{ruledtabular}
  \begin{tabular}{llll}
     & Single-gate planar & Dual-gate & Symmetric GAA \\
    \colrule
    Gate arrangement & one-sided & top and bottom & surrounding \\
    Potential symmetry & asymmetric & selectable by bias & symmetric (ideal) \\
    Common-mode potential & present & independently tunable & present \\
    Interlayer differential potential & generally finite & independently tunable & ${\approx}0$ (ideal) \\
    Carrier-density control & yes & strong, independent & strong \\
    Displacement field & mixed with density & generated/suppressed at will & suppressed to leading order \\
    Interlayer Stark effect & possible & common: off / diff: on & suppressed \\
    $\Delta_{K\Gamma}$ under gating & may shift with $V_\mathrm{GS}$ & tuned separately from density & ${\approx}$ fixed \\
    $K$-point PT degeneracy & generally lifted & controlled by differential bias & preserved to leading order \\
  \end{tabular}
  \end{ruledtabular}
\end{table*}

Which mechanism operates depends on the gate geometry
(Table~\ref{tab:gaa_planar}). In a single-gate planar device the vertical
field is applied predominantly from one side of the channel, so the
potential is asymmetric about the channel midplane and a finite
antisymmetric interlayer field mixes carrier-density control with Stark
modification of $\Delta_{K\Gamma}$.
A dual-gate device is more general:
decomposing the top- and bottom-gate voltages, for matched top- and
bottom-gate capacitances, into
$V_\mathrm{cm}=(V_\mathrm{top}+V_\mathrm{bot})/2$ and
$V_\mathrm{diff}=(V_\mathrm{top}-V_\mathrm{bot})/2$, the common mode
controls the carrier density while the differential mode generates the
antisymmetric displacement field, so that occupation control and
band-structure control can be exercised \emph{independently}. This makes
the dual-gate geometry a natural experimental test of the present model:
sweeping $V_\mathrm{cm}$ at fixed $V_\mathrm{diff}$ probes the
Fermi-filling redistribution at fixed $\Delta_{K\Gamma}$, while sweeping
$V_\mathrm{diff}$ probes the field-induced change of $\Delta_{K\Gamma}$.
Such displacement-field control of the $\Gamma$--$K$ balance has already
been demonstrated in dual-gated few-layer
WSe$_2$\cite{Movva2018_valleyPRL,Lai2023_valleytronic,Wei2025_twistedWSe2KG}.
In an ideally symmetric gate-all-around device, by contrast, the gate
surrounds the channel and the electrostatic potential is symmetric about
the midplane, $\phi(+z)=\phi(-z)$, where $z$ is measured from that
midplane along the stacking direction.
The two layers are then
electrostatically equivalent, the antisymmetric interlayer field vanishes
to leading order, and the gate acts almost purely as a common-mode
carrier-density control. The symmetric GAA thus structurally realizes the
common-mode limit without requiring independent balancing of two gate
voltages, and treating $\Delta_{K\Gamma}$ as gate-independent in
Sec.~\ref{sec:results} is the appropriate leading-order description in
this limit.

To fix the magnitude and sign of the Stark term where the layer symmetry
\emph{is} broken---by single- or top-gated operation, a finite dual-gate
displacement field, an asymmetric dielectric environment, or ferroelectric
gating\cite{Muller2011APL,Lin2025NatElectron}---we performed DFT+SOC
calculations of the unstrained bilayer under a uniform out-of-plane
(displacement) field (Supplemental Material~\cite{supplemental}).
The
field \emph{increases} $\Delta_{K\Gamma}$---from $27$~meV at zero field to
$\approx 0.15$~eV at $0.5$~V\,nm$^{-1}$---and does so for both field
polarities. The near-even dependence follows because field reversal
exchanges the two layer-polarized $K$ branches while, by our convention,
$\Delta_{K\Gamma}$ is read from the \emph{upper} branch, so the extracted
splitting rises for either sign. A full attribution awaits the
layer-projected analysis left to future work.
Thus a symmetry-breaking
vertical field drives the system \emph{away} from, rather than toward, the
near-degenerate regime, and cannot be used to reach it from the present
positive-$\Delta_{K\Gamma}$ starting point. (These calculations are for
the unstrained 2H bilayer with $K$ at the valence-band maximum, evaluated
with PBE-D3(BJ) and HSE06 over the field range studied; a structure that
starts on the $\Gamma$-dominated side, $\Delta_{K\Gamma}<0$, could instead
be driven \emph{toward} the crossover by the same field.) This direction
is consistent with the experimental transfer of holes from $\Gamma$ to $K$
under a transverse field in \emph{trilayer}
WSe$_2$\cite{Movva2018_valleyPRL}, although the layer number, temperature,
and carrier-density regimes differ from the present bilayer calculation.
The same symmetry-breaking field also lifts the PT doublet at $K$
(Table~\ref{tab:gaa_planar}), so the fixed-$g_K=4$ two-valley model
applies strictly in the layer-symmetric regime; a branch-resolved
treatment of the two $K$ states would be required at finite displacement
field.

The design implication is straightforward. Because a symmetry-breaking
field only \emph{raises} $\Delta_{K\Gamma}$, it is the \emph{symmetric}
wrap-around gate---which suppresses the antisymmetric interlayer field and
leaves the gate to control carrier density alone---that avoids an
additional field-induced increase of $\Delta_{K\Gamma}$ and so preserves
the zero-displacement-field valley landscape on which the
valley-redistribution mechanism depends. Strain, not the vertical field,
is the parameter that tunes $\Delta_{K\Gamma}$ toward or away from
degeneracy. A quantitative first-principles treatment of the
asymmetric-field regime, including the layer-resolved character of the
Stark response, is left to future work.

\subsection{Robustness of the Valley Splitting}
\label{sec:robustness}

The bilayer splitting itself deserves comment.
Additional DFT+SOC calculations (Supplemental
Material~\cite{supplemental}) show that $\Delta_{K\Gamma}$ is numerically
well converged with respect to $k$-point sampling and vacuum spacing, but
is strongly sensitive to the physical approximations.
It varies by more
than $300$~meV across van der Waals functionals (with DFT-D2 and optB88
even placing $\Gamma$ at the valence-band maximum), and the hybrid HSE06
functional gives $\approx0.1$~eV against the PBE value of $27$~meV.
This sensitivity is largely traceable to the interlayer W--W distance,
which the various functionals predict over a wide range and which
controls the interlayer hybridization at $\Gamma$.
A dedicated
first-principles study finds this distance to be the single dominant
determinant of the splitting and shows that hydrostatic or uniaxial
pressure tunes it continuously, driving the valence-band maximum from
$K$ to $\Gamma$ at $0.5$--$3$~GPa\cite{Brzezinska2025_pressure}.
The interlayer stacking is a further degree of freedom---the 2H and 3R
registries are nearly degenerate in energy and modify the band
structure~\cite{He2014_stacking}, and chemical-vapor-deposited bilayer
WSe$_2$ is found to contain coexisting 2H and 3R
domains~\cite{DiBerardino2025_2H3R}.

\begin{figure*}[tb]
  \includegraphics[width=\textwidth]{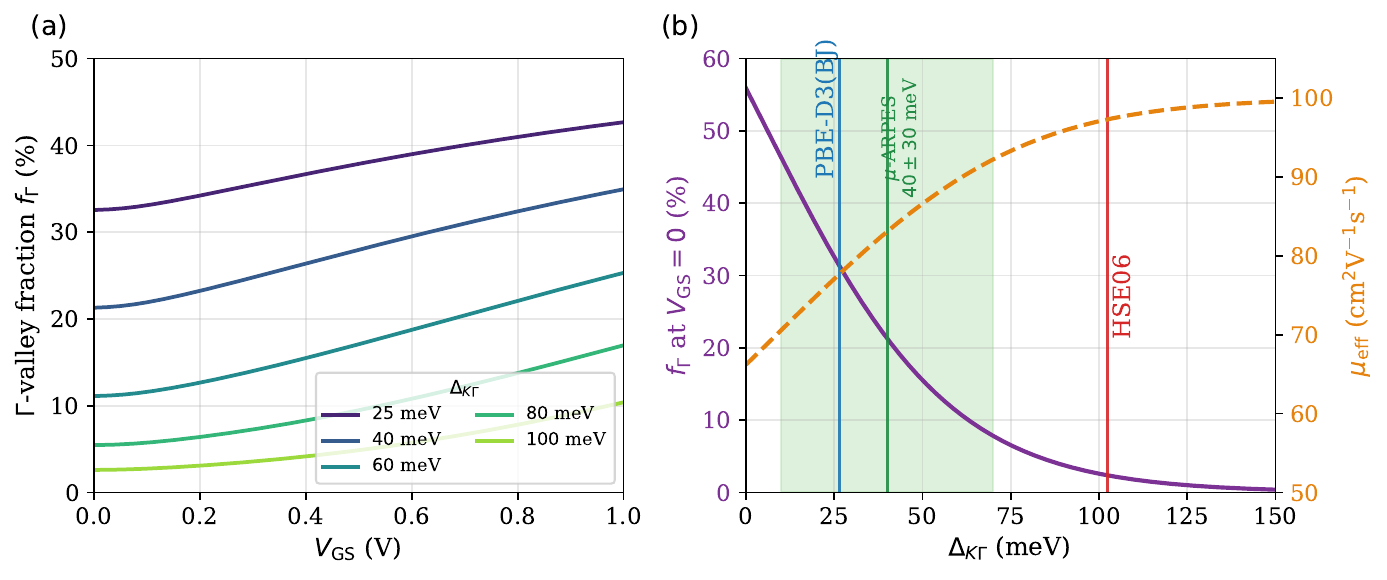}
  \caption{Device response across the $\Delta_{K\Gamma}$ landscape, with
    $\Delta_{K\Gamma}$ treated as a free parameter in the two-valley
    model (unstrained-bilayer effective masses).
    (a)~$\Gamma$-valley hole fraction $f_\Gamma$ versus $V_\mathrm{GS}$
    for $\Delta_{K\Gamma} = 25, 40, 60, 80, 100$~meV.
    (b)~$f_\Gamma$ at $V_\mathrm{GS}=0$ (left axis) and
    $\mu_\mathrm{eff}$ at $V_\mathrm{GS}=0.5$~V (right axis) versus
    $\Delta_{K\Gamma}$; $V_\mathrm{GS}=0$ characterizes the equilibrium
    valley mixture, whereas $V_\mathrm{GS}=0.5$~V represents the on-state
    mobility.
    Vertical lines mark the PBE-D3(BJ) ($26.5$~meV) and HSE06 ($102$~meV)
    values computed in this work (Supplemental
    Material~\cite{supplemental}); the green line and shaded band mark the
    experimental $\mu$-ARPES value $0.04\pm0.03$~eV for the 2H bilayer
    \cite{Wilson2017_muARPES}.
    The valley redistribution is strongest at small $\Delta_{K\Gamma}$
    and fades toward single-valley $K$ transport as $\Delta_{K\Gamma}$
    grows.}
  \label{fig:dkglandscape}
\end{figure*}

The splitting is therefore best viewed not as a fixed material constant
but as a device- and structure-dependent \emph{landscape} parameter set
by stacking, strain, pressure, and the dielectric and electrostatic
environment.
Theoretical
estimates additionally depend on the exchange--correlation functional.
Crucially, the strain response is consistent across functionals, so the
valley-redistribution mechanism is preserved.
What changes is where on
the landscape a given sample sits.
The device consequences of moving along this landscape are shown in
Fig.~\ref{fig:dkglandscape}, which treats $\Delta_{K\Gamma}$ as a free
parameter in the two-valley model.
Figure~\ref{fig:dkglandscape}(a) gives the $\Gamma$-valley occupation
$f_\Gamma(V_\mathrm{GS})$ for $\Delta_{K\Gamma}$ from $25$ to $100$~meV,
and Fig.~\ref{fig:dkglandscape}(b) gives $f_\Gamma$ at $V_\mathrm{GS}=0$
and $\mu_\mathrm{eff}$ at $V_\mathrm{GS}=0.5$~V versus $\Delta_{K\Gamma}$,
with the PBE-D3(BJ) and HSE06 values of this work marked.
Near the PBE value both valleys remain substantially populated
($f_\Gamma\approx31\%$) and appreciable gate tunability of
$\mu_\mathrm{eff}$ survives (the valley susceptibility peaks at the
smaller value $\Delta_{K\Gamma}\approx k_BT\ln r_N \approx 6$~meV).
By the HSE06 value the $\Gamma$ valley is largely depopulated
($f_\Gamma\approx2\%$) and the channel is nearly single-valley.
The valley redistribution is thus strongest at small $\Delta_{K\Gamma}$
and fades toward single-valley $K$ transport as the splitting grows.

Experimental and theoretical estimates place bilayer WSe$_2$ near or
across the boundary between the mixed-valley and predominantly
$K$-valley regimes.
Samples at the lower end of the reported
$\Delta_{K\Gamma}$ range (e.g.\ the $\mu$-ARPES value) show strong
two-valley occupation, whereas larger splittings (e.g.\ HSE06) approach
the single-$K$-valley limit. Where a given sample lands on the landscape
is set by the structure-, stacking-, pressure-, and strain-dependent
value of $\Delta_{K\Gamma}$.
Within the regime, the strain direction then selects the objective.
Because tensile strain reduces $\Delta_{K\Gamma}$ while compressive strain
increases it, moving \emph{toward} the crossover (tensile strain,
stacking, or dielectric engineering) maximizes the \emph{sensitivity} of
$\mu_\mathrm{eff}$ to gate bias and strain, whereas moving \emph{away}
from it (compressive strain) depopulates the heavy $\Gamma$ valley and
maximizes the effective mobility and hence the on-current
[Sec.~\ref{sec:results}, Fig.~\ref{fig:id_combined}].
These are complementary uses of the same valley-redistribution
mechanism---tunability versus on-state mobility---rather than
competing prescriptions.

\subsection{Model Limitations}
\label{sec:limits}

Several approximations in the present model deserve mention.

First, the ratio $\mu_\Gamma/\mu_K = m^*_K/m^*_\Gamma$ assumes a
common intravalley scattering time for both
valleys\cite{Fivaz1967}.
This is a deliberate minimal choice rather than a claim about the
microscopic scattering physics.
The two valleys differ in orbital character---the $K$ states derive
from in-plane W $d_{x^2-y^2}/d_{xy}$ orbitals confined within a single
layer, whereas the $\Gamma$ states are built from interlayer-hybridized
$d_{z^2}$ and Se $p_z$ orbitals---so their deformation potentials
and phonon couplings are not identical\cite{Jin2014}.
A first-principles evaluation of the electron--phonon-limited mobility
of each valley is beyond the scope of this work.
The equal-$\tau$ assumption is adopted because it introduces no
additional free parameter and because it yields only the weak inequality
actually required, $\mu_\Gamma < \mu_K$. The DFT effective masses
($m^*_\Gamma > m^*_K$) suggest this inequality under comparable
scattering times but do not by themselves prove it. The qualitative results
require only the physically motivated assumption $\mu_\Gamma < \mu_K$,
not the specific equal-$\tau$ estimate (verified by the
$\mu_\Gamma/\mu_K$ sensitivity scan in the Supplemental
Material~\cite{supplemental}).
The qualitative trend---increasing $\Gamma$-valley occupation
monotonically reduces $\mu_\mathrm{eff}$---holds for \emph{any}
$\mu_\Gamma/\mu_K < 1$, independently of $\tau$.
This is verified explicitly in the Supplemental
Material~\cite{supplemental}, where $\mu_\Gamma/\mu_K$ is varied over
the range $0.2$--$0.8$ and all reported trends are unchanged.
The absolute values of $\mu_\mathrm{eff}$ in Sec.~\ref{sec:results}
should therefore be interpreted as model-specific estimates that scale
with $\mu_K$, rather than as quantitative predictions.

Second, intervalley scattering between $K$ and $\Gamma$ is neglected.
At $K$ the occupied states form the upper spin--layer doublet, which lies
$467$--$482$~meV above the lower doublet. This large separation suppresses
scattering between the upper and lower manifolds (it does not concern
scattering \emph{within} the degenerate upper doublet). The argument does
not extend to the $\Gamma$ valley, which carries no comparable manifold
splitting because $\Gamma$ is a time-reversal-invariant momentum.
$K$--$\Gamma$ scattering is instead suppressed by the large momentum
transfer required, which restricts it to zone-edge phonons, and by the
weak overlap between the in-plane and interlayer-bonding orbital
characters noted above.
Near the K--$\Gamma$ degeneracy, however, the available phase space
grows, so this channel would act as an additional mobility-reduction
mechanism in the $\Gamma$-dominated
regime\cite{Schaibley2016_valleytronics,Xu2014_spinpseudo}.
Including it is expected to modify, and plausibly enhance, the reduction
of $\mu_\mathrm{eff}$ with increasing $\Gamma$ occupation near the
crossover; its quantitative magnitude, and even its detailed gate
dependence, would however require a microscopic calculation of the
intervalley rates (which depend on the initial valley, the final-state
density of states, the phonon energies, deformation potentials,
occupation, and Pauli blocking) and are beyond the scope of the present
model.

Third, contact resistance is not included in the present model.
In 2D-material FETs, contact resistance can be a significant fraction
of the total resistance\cite{Allain2015,Das2013}. As a gate-, strain-,
and layer-independent series resistance it would primarily reduce the
absolute current level and would not affect the valley-driven trends
identified here. A Schottky-barrier contact whose resistance itself
depends on gate bias, strain, or band alignment is beyond the present
scope.
Fourth, $\Delta_{K\Gamma}$ is treated as independent of gate bias.
As shown in Sec.~\ref{sec:discussion_efield}, this is the correct
leading-order description for a symmetric wrap-around gate, for which
the antisymmetric interlayer field vanishes to leading order and the
gate acts as a common-mode carrier-density control.
The assumption would have to be relaxed for asymmetrically gated
geometries, where an out-of-plane field increases $\Delta_{K\Gamma}$
(Sec.~\ref{sec:discussion_efield}).

Fifth, the strain comparison is performed at \emph{fixed} total
equilibrium hole density $p_0$, which isolates the effect of the
relative K--$\Gamma$ landscape. Because the DFT band energies are
referenced within each structure to its own valence-band maximum, the
strain-induced shifts of the \emph{absolute} band alignment relative to
the vacuum level---and hence of the electron affinity, work-function
alignment, threshold voltage, and contact barrier---are not included.
The reported $I_\mathrm{on}(\varepsilon)$ therefore represents the change
in the intrinsic valley-averaged transport at a common carrier density,
rather than an absolute prediction of the current at a fixed external
$V_\mathrm{GS}$ for a single device placed under strain.

\subsection{Compressive Strain in Realistic Device Environments}

While tensile strain can be readily applied using flexible substrates
or bending techniques,\cite{He2013,Conley2013} the realization of
uniform in-plane compressive strain in atomically thin materials is
generally more challenging due to buckling or wrinkle
formation\cite{Johari2012,Scalise2014}.
However, compressive strain can arise naturally from thermal expansion
mismatch between the 2D channel and surrounding dielectrics during
fabrication or operation\cite{Desai2016}.
In addition, in GAA architectures, the channel is fully encapsulated
by high-$\kappa$ materials, which mechanically constrain the lattice
and may help stabilize compressive strain.
Therefore, the compressive-strain regime explored in this work should
be regarded as experimentally relevant, particularly in fully
encapsulated or integrated device structures.
Both ends of the useful strain range are thus accessible: higher
on-state mobility (compressive strain, larger $\mu_\mathrm{eff}$ and
$I_\mathrm{on}$) through encapsulation and thermal-mismatch strain, and
maximal valley tunability (tensile strain, toward the crossover) through
flexible-substrate bending. A device can therefore be positioned along
the $\Delta_{K\Gamma}$ landscape according to whether on-state mobility
or valley tunability is the target, with device-induced compressive
strain being the more natural fit for encapsulated GAA geometries.

\section{Conclusion}

A combined DFT and analytical model study of hole transport in
WSe$_2$ GAA FETs has been presented.
The central finding is that the bilayer K--$\Gamma$ splitting
$\Delta_{K\Gamma}$ is not a universal material constant but a
device- and structure-dependent landscape parameter: while it is
numerically converged, its value spans from $27$~meV (PBE) to
$\approx 0.1$~eV (HSE06), moves by more than $300$~meV across van der
Waals functionals, and depends on the interlayer stacking.
We therefore treat $\Delta_{K\Gamma}$ as a free parameter and map how it
controls the device physics.
An analytical two-valley model then predicts the following.
(i)~Because the quantum capacitance stays well below the oxide
capacitance deep below threshold, the minimum subthreshold swing
stays near the thermionic limit of $60$~mV~dec$^{-1}$ for every layer
number.
(ii)~The effective mobility is controlled by the K-to-$\Gamma$
occupation ratio, and the valley-redistribution mechanism is strongest
when $\Delta_{K\Gamma}$ is small (comparable to $k_BT$), fading toward
single-valley $K$ transport as the splitting grows.
(iii)~In this small-splitting regime, biaxial strain tunes
$\mu_\mathrm{eff}$ and the on-current through the valley population while
SS is preserved at the thermionic limit. The off-state current scales
with the same $\mu_\mathrm{eff}$, so the on/off ratio stays nearly
strain-independent. The device-relevant outcome is thus a decoupling of
mobility control from the subthreshold swing, distinct from strategies
that tune mobility primarily by modifying the scattering rate.

The design rule has two facets: the small-$\Delta_{K\Gamma}$ regime
maximizes the valley \emph{tunability}---small changes in strain or
carrier density then produce large shifts of the hole population between
the $K$ and $\Gamma$ valleys, and hence of $\mu_\mathrm{eff}$---whereas
larger $\Delta_{K\Gamma}$ (compressive strain) depopulates the heavy
$\Gamma$ valley and maximizes the on-state \emph{mobility}.
Because tensile strain lowers $\Delta_{K\Gamma}$ while compressive strain
raises it, strain---together with stacking and the dielectric
environment---is the control that positions a device on this landscape.
In a symmetric gate-all-around device the gate controls carrier density
at essentially fixed splitting, because the antisymmetric interlayer
field that would otherwise modify $\Delta_{K\Gamma}$ through the Stark
effect vanishes to leading order in the ideal midplane-symmetric limit.
DFT+SOC calculations show that a symmetry-breaking out-of-plane field
instead \emph{increases}
$\Delta_{K\Gamma}$, so that preserving layer symmetry is itself a design
requirement for avoiding a displacement-field-induced departure from the
operative regime.
Beyond its well-known electrostatic advantages, the symmetric
gate-all-around geometry structurally favors the common-mode limit and
suppresses the antisymmetric interlayer potential
(Table~\ref{tab:gaa_planar}), without requiring independent balancing of
top- and bottom-gate voltages.
More broadly, this work identifies valley population as an intrinsic and
gate-accessible design parameter for 2D transistors---one with no direct
counterpart in conventional single-valley semiconductors---and
identifies bilayer WSe$_2$ as a structurally tunable platform whose
transport response is set by its realized $\Delta_{K\Gamma}$.

\begin{acknowledgments}
This work was supported by JSPS KAKENHI (Grants No.~JP25K01609,
No.~JP22H05473, and No.~JP21H01019) and JST CREST (Grant
No.~JPMJCR19T1). K.W.\ acknowledges financial support for Basic
Science Research Projects (Grant No.~2401203) from the Sumitomo
Foundation.
\end{acknowledgments}

\bibliography{references}

\end{document}